\documentclass[preprint,prm]{revtex4-2}
\usepackage{geometry}           
\usepackage{graphicx}
\usepackage{subfig}
\usepackage{amsmath,amssymb}
\graphicspath{{./}}
\usepackage{color}
\usepackage[normalem]{ulem}

\newlength{\figwidth}
\newcommand{\fref}[1]{Fig.\,\ref{#1}}

\newcommand{\eref}[1]{Eq.\,(\ref{#1})}

\newcommand{\sref}[1]{Sec.\!~\ref{#1}}

\newcommand{\cref}[1]{Ref.\,\cite{#1}}

\newcommand{\vs}{{\it vs.}\! }
\newcommand{\ie}{{\it i.e.}\!\, }

\newcommand{\etc}{{\it etc.}\! }
\newcommand{\etal}{{\it et al.} }

\newcommand{\abinitio}{{\it ab initio} }

\newif\ifclean
\cleantrue  

\begin{document}
\title{\bf Tuning the critical Li intercalation concentrations for MoX$_2$ bilayer phase transitions}
\author{C.D. Spataru}\email{cdspata@sandia.gov}
\affiliation{Sandia National Laboratories, Livermore, CA 94551}
\author{M.D. Witman}
\affiliation{Sandia National Laboratories, Livermore, CA 94551}
\author{R.E. Jones}
\affiliation{Sandia National Laboratories, Livermore, CA 94551}


\begin{abstract}
Transition metal dichalcogenides (TMDs), such as MoS$_2$, are known to undergo a structural phase transformation as well as a change in the electronic conductivity upon Li intercalation.
These properties make them candidates for charge-tunable ion-insertion materials  that could be used in electro-chemical devices. 
In this work we study the phase stability and electronic structure of H and T' Li-intercalated MoX$_2$ bilayers with X=S, Se, or Te.
Using first-principles calculations in combination with classical and machine learning modeling approaches, we find that the H phase is more stable at low Li concentration for all X, and the critical Li concentration at which the T’$\to$H transition occurs decreases with increasing mass of X. 
Furthermore the relative free energy of the two phases becomes less sensitive to Li insertion with increasing atomic mass of the chalcogen atom X.
While the electronic conductivity increases with increasing ion concentration at low concentrations, we do not observe a (positive) conductivity jump at the phase transition from H to T'.
\end{abstract}

\maketitle

\clearpage

\section{Introduction}

Autonomous vehicles (AVs) will serve global energy security goals as they offer efficiency opportunities that could enable significant reduction in fuel consumption. 
Unfortunately, with the current technology the electrical power associated with on-board sensing and calculations for driving decisions could reach quite large values of the order of kW \cite{murray2018basic}.
This is a consequence of the fact that silicon technology uses about a pJ/operation \cite{fuller2017li}. 

Neuromorphic computing is a novel computing architecture that holds the promise for advanced computing with high energy efficiency, while being well suited for artificial intelligence (AI) tasks.
Neuromorphic computing could have a profound impact on computation for autonomous vehicles and other sectors that require edge computing.
One route for realizing novel devices for neuromorphic computing is to leverage ionic/electronic transport properties in electro-chemical devices.
However, a major roadblock in the development of electro-chemical devices is the lack of understanding of the fundamental mechanisms that lead to conductivity changes in the ion-insertion materials used in neuromorphic devices.

Recently, a novel electro-chemical device called a Li-Ion Synaptic Transistor for Low Power Analog Computing (LISTA) has been pioneered by Talin \etal~\cite{fuller2017li}. 
LISTA is 3-terminal redox transistor which allows ion-insertion into a channel by applying a small gate voltage.
Through reversible Li intercalation, the channel electronic conductance (synaptic weight) can be gradually changed. Hundreds of conducting states can be modulated in a controllable fashion with switching times ($< \mu$s)  comparable to those measured in biological synapses \cite{fuller2019parallel,yang2018all} and much lower power ($<$ fJ/operation \cite{fuller2017li}) than Si-based CMOS transistors.

In order to take the proof-of-concept LISTA device to a scalable technology, it is important to understand how it works at a fundamental level.  
Despite progress on the experimental front, the mechanisms that lead to conductivity changes are not fully understood.
In particular it is not clear how the structural and electronic properties of the ion-insertion material evolve as a function of ion concentration.  
For example, do ions intercalate into the channel in an ideal fashion, {\it i.e} gradually and with uniform concentration, or rather do they form co-existing domains of insulating and metallic phases with sizes that depend on gate voltage? 
This and related questions regarding the equilibrium and nonequilibrium intercalation regimes can be addressed computationally via mesoscopic modeling approaches \cite{nadkarni2019modeling}. 
However, such models require knowledge of the phase stability diagram as well as the electronic properties of the ion-insertion materials as function of (uniform) ion concentration. 
Such knowledge can be obtained from first-principles studies.

From a materials point of view, quasi-two dimensional layered systems are attractive due to their layered structure. 
Indeed, they can be very useful in electrochemical devices because the weak van der Waals interaction between layers facilitates ion intercalation. 
As such, layered systems can find applicability in electrode materials for batteries with high energy storage, for example.
Transition metal dichalcogenides (TMDs), {\it i.e.} MoS$_2$, MoSe$_2$ or MoTe$_2$, represent a particular class of layered materials.
These materials undergo a structural phase transformation upon ion intercalation. 
This transition is accompanied by a change in the electronic conductivity, which makes TMDs candidates for charge tunable ion-insertion materials in electro-chemical devices for neuromorphic computing.
Moreover, two-dimensional, few-layer TMDs are of interest for small-scale device fabrication which is desirable since the write energy per operation (resistance) increases with channel size \cite{fuller2017li}. 
While there is a large body work on ion intercalation in bulk TMDs \cite{enyashin2013line,benavente2002intercalation,Zhao} and ion adsorption on monolayer TMDs \cite{sun2016origin}, fewer studies so far have focused on few-layer TMDs \cite{zhang2018reversible} such as bilayers \cite{pandey2016phase,lu2020lithium}.
Of the few studies of bilayers, Pandey \etal \cite{pandey2016phase} examined the phase transition of MoS$_2$ induced by lithiation to a heterostructure involving a  metallic phase.
The later work of Lu \etal \cite{lu2020lithium} also used \abinitio simulations to examine the relative rotation of the bilayers as a means of controlling the band structure of Li intercalated MoS$_2$.

In this work we perform a first-principles based study of the phase stability and electronic structure evolution of Li ion-intercalated MoS$_2$, MoSe$_2$, MoS$_2$ bilayer TMDs.
Changes in phase and the attendant changes in electronic structure in MoX$_2$ bilayers have ramifications on the performance of neuromorphic and related devices utilizing these materials.
We hypothesize that \abinitio Density Functional Theory (DFT) can be integrated with machine learning to efficiently predict the phase stability and electronic structure evolution of ion-insertion layered materials.
We explore this path, together with the more established cluster expansion method, to obtain thermodynamic properties, the phase stability diagram, and related electronic properties.

\section{Theoretical framework}
\begin{figure}
\centering
{\includegraphics[width=\textwidth]{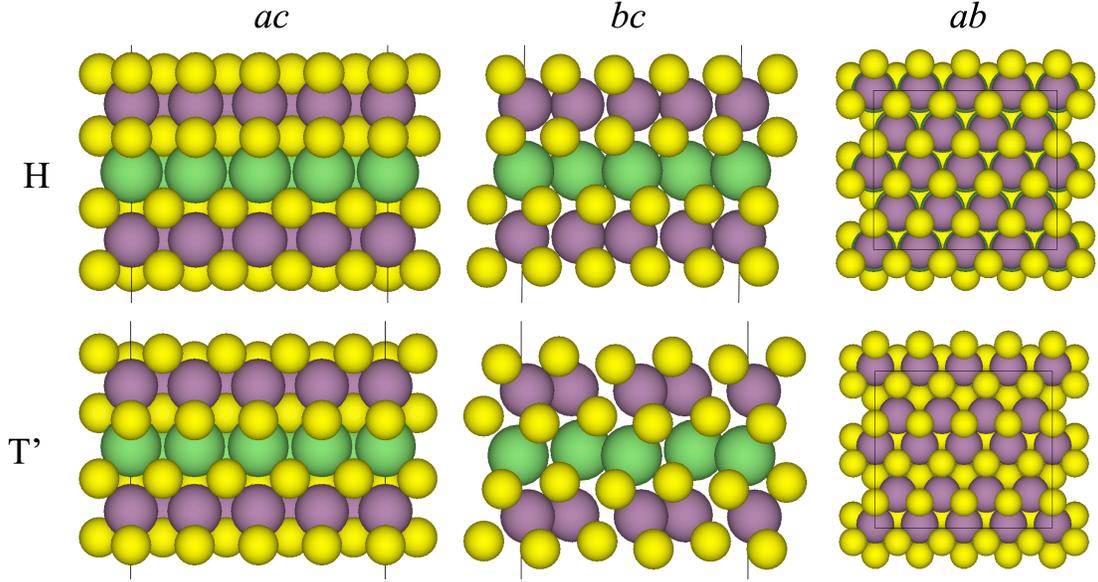}}
\caption{Atomic structure of bilayer MoX$_2$ layered material (top panels) H
  phase and (bottom panels) T' phase. Left and middle panels: side views of the bilayer. Right panels: top view. Color code for the atoms: purple=Mo, yellow=X, green=Li.
}
\label{fig:structure}
\end{figure}

TMDs may take several polytype forms, such as the semiconducting H phase, metallic T phase, and semimetallic T' phase.
In few-layer TMDs, the metallic T phase is unstable due to a Peierls distortion that dimerizes the metal atoms leading to the formation of the T' phase \cite{li2016ferroelasticity}.
Thus, we consider only two competing phases, namely H and T' which are illustrated in \fref{fig:structure}.
A large fraction of our computational effort consists on estimating the free energy of these competing phases, from which a phase diagram as a function of Li concentration will be constructed.
Ultimately, this allows us to make predictions regarding the electronic conductivity of Li ion-intercalated bilayer TMDs as function of ion concentration.

\subsection{Free energy and phase stability} \label{sec:free_energy}
To motivate our methodology a brief review of fundamental statistical mechanics is needed.
The canonical, Helmholtz free energy is
\begin{equation} \label{eq:F}
F(N_\text{Li},T)  = k_B T \ln(Z(N_\text{Li},T))
\end{equation}
where $k_B$ is Boltzmann's constant, $T$ is temperature, and  
\begin{equation} \label{eq:zNT}
Z(N_\text{Li},T) = \sum_i \exp\left( -\frac{E_i}{k_B T} \right) 
\end{equation}
is the partition function. 
In \eref{eq:zNT} the sum is over all possible configurations with $N_\text{Li}$ lithium atoms, and the energy $E_i$ is energy of configuration $\chi_i$ relative to a chosen reference state.
The ensemble average of a quantity $X$ can be computed via
\begin{equation} \label{eq:ensemble_average}
\langle X \rangle_{N_\text{Li}\text{T}} 
= \frac{1}{Z}  \sum_i X_i \exp\left( \frac{E_i}{k_B T} \right) 
\end{equation}
For brevity we have suppressed the dependence of the free energy on the system stress as we focus on fully relaxed systems, \ie at zero stress; and, as a consequence,  we will refer to this ensemble as the NT ensemble.
A minimum in $F(N_\text{Li},T)$ indicates the most likely concentration of Li at a given $T$.
The difference between phases 
\begin{equation} \label{eq:dF}
\Delta F(N_\text{Li},T) \equiv F(N_\text{Li},T)_\text{T'}-F(N_\text{Li},T)_\text{H}  
=  k_B T \ln \left( \frac
{ \left( \sum_i \exp\left( -\frac{E_i}{k_B T} \right) \right)_\text{T'} }
{ \left( \sum_i \exp\left( -\frac{E_i}{k_B T} \right) \right)_\text{H} } 
\right)
\end{equation}
indicates which phase is more stable and eliminates the dependence on a reference state.

The grand canonical free energy $\Phi$ takes a similar form to $F$ in \eref{eq:F}
\begin{equation}
\Phi(\mu,T) = k_B T \ln(\Omega(\mu,T))
\end{equation}
where the partition function is 
\begin{equation} \label{eq:zmuT}
\Omega(\mu,T) = \sum_i \exp\left( \frac{\mu N_i}{k_B T} \right) Z(N_i,T) \ .
\end{equation}
In this case the sum is over all possible configurations from no Li, $N_\text{Li}=0$, up to $N_\text{Li}=N_\text{sites}$ at the saturation of all low energy sites.
Again we employ relaxed systems and refer to this as the $\mu$T ensemble.
In the present context of intercalation we introduce the site occupancy fraction  $\phi = N_\text{Li}/N_\text{sites}$ and the formation energy \cite{van1998first}
\begin{equation} \label{eq:formE}
\tilde{E}(\phi) = E(\phi) - \phi \left[E(1)-E(0)\right] -  E(0)
\end{equation}
The grand canonical potential provides information regarding how  many Li reside on/in the bilayer when in equilibrium with an environmental reservoir of lithium (as in an electrochemical device) at chemical potential $\mu$.
With the free energy $\Phi$ the average Li site occupancy can be computed: 
\begin{equation}
\langle \phi \rangle_{\mu\text{T}} = - \frac{1}{N_\text{sites}} \frac{\partial \Phi}{\partial \mu} 
=  \frac{1}{N_\text{sites}} \sum_{i=0}^{N_\text{sites}} N_i \frac{\exp\left( \frac{\mu N_i}{k_B T} \right) Z(N_i,T)}{\Omega(\mu,T)}
\end{equation}

In these free energy expressions the contribution to the entropy comes entirely from the configurational degrees of freedom.
We omit dependence on vibrational modes, electronic modes, \etc
While the vibrational entropy can be significant, its contribution to the free energy difference between phases is expected to be minor;  the variation of the vibrational entropy across phases of the same chemical species is typically small  so that it does not affect significantly the relative phase stability \cite{urban2016computational,van1998first}.
Similarly, while the configurational entropy may converge slowly with system size, its variation between chemically similar phases is expected to converge fast enough to make it tractable to compute with feasible DFT system/supercell sizes.
We explore the validity of this assumption together with further details on approximating the ensemble average in Appendix A.

\subsection{Method for obtaining total energies}
To calculate the relative free energies and thus the phase diagram we
need a set of energies $\{ E_i \}$ that sample the configurational space of each phase.
We obtain these from \abinitio density functional theory (DFT) calculations and hence are limited to small systems with a fixed number of bilayer lattice atoms and varying number of intercalated lithium atoms.
The configurational space of these small systems can be sampled exhaustively using a surrogate model of the energies and other quantities of interest as a function of the bilayer configuration.
We explored a number of surrogate models including classical cluster expansion and a variety of machine learning methods.
   
\subsubsection{{\it Ab initio} calculations}

We calculate total energies as well as the electronic density of states (DOS) of Li ion-intercalated bilayer TMD structures within the framework of \abinitio DFT using the VASP code \cite{kresse1996efficiency}.
We use projector augmented wave (PAW) pseudo-potentials \cite{blochl1994projector} and treat the exchange and correlation terms within the generalized gradient approximation with the PBE parameterization \cite{perdew1996generalized}.
The energy cutoff was set to 400 eV.

We consider bilayer structures where the two monolayers are rotated by 180$^\circ$ with respect to each other.
The translational alignment between the two MoX$_2$ monolayers is such that the Mo atoms are vertically aligned (refer to \fref{fig:structure}), since within our DFT parameter settings we find the energy for this configuration to be lower than for other alignments in the presence of Li-ion intercalation. In particular, this translational alignment yields for MoS$_2$ a total energy (per a rectangular $4\times2$ supercell that contains $1$ intercalated Li ion) $22$ meV lower than the corresponding lowest energy of the intercalated AA’ stacking (where the top layer’s Mo atoms are aligned with the bottom layer’s S atoms and viceversa \cite{he2014,lu2020lithium}).

We construct systems using supercells obtained by laterally replicating the rectangular unit cell of bilayer MoX$_2$ with X=S, Se or Te.
We use $4\times2$ lateral replicas with lateral area of $\approx$ 1.4$\times$1.2 nm$^2$ and vacuum separation along the out-of-plane direction of more than $15$ \AA .
There are 32 Mo atoms and 64 X atoms in our supercells in the absence of ion intercalation, and $N_\text{sites}=16$ lithium atoms at saturation ($\phi=1$) of the low energy sites.

\begin{figure}
\centering
{\includegraphics[width=\textwidth]{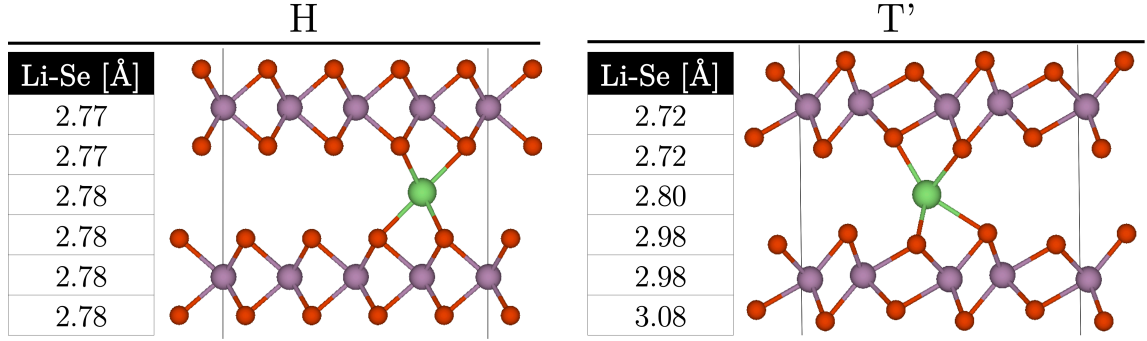}}
\caption{Side view of the atomic structure of bilayer MoSe$_2$ intercalated with a single Li ion. (a) H phase. (b) T' phase. Color code for atoms: red=Se; purple=Mo, green=Li.
The tables give the bond lengths from the Li to its six closest Se neighbors for each phase.
}
\label{fig:Tp coordination}
\end{figure}

Even with only tens of sites, there is a large configurational space associated with Li ion intercalation in between the MoX$_2$ monolayers of the bilayer.
To explore the configurational space in order to compute ensemble averages, it is sufficient to sample only stable configurations as they carry most of the weight in \eref{eq:ensemble_average} given that they correspond to local energy minima.
For each phase we identify low energy ion-intercalation sites and select the most stable one, as we find that this crystallographic location has an energy several hundred meV lower than the other, metastable configurations.
In the case of the H phase we find that the Li ion is most stable when positioned half-way in between two vertically aligned Mo atoms from different monolayers (see \fref{fig:Tp coordination}a)) \cite{lu2020lithium}.
This results in octahedral coordination to the X atoms (with bond length of about $2.8$ \AA).
In the T' case the most stable Li site is not vertically aligned with Mo while coordination with the closest six X neighbors is distorted from the ideal octahedral case with four different bond lengths ranging from $2.72$ \AA\  to $3.06$ \AA\ (see \fref{fig:Tp coordination}b).
We also checked in the MoS$_2$ case for both phases that Li ion adsorption yields an energy significantly larger (by more than $100$ meV) than Li ion intercalation, hence the focus of this work is on ion intercalation.

The Li ion-intercalation sites in the 4$\times$2 supercell form a 16-site layer sublattice, \ie there is one possible Li intercalation site for each Mo$_2$X$_4$ bilayer formula unit. 
The Li-concentration $x$ in the stoichiometric formula  Li$_x$MoX$_2$ ranges from $x=0$ (no intercalation, $N_{Li}=0$, $\phi=0$) to $x=0.5$ at full ($N_\text{Li}= 16$, $\phi=1$) Li sublattice occupancy.
The number of possible configurations for fixed $N_\text{Li}$ is $N_\text{sites} ! / ( N_\text{Li}!  (N_\text{sites}  - N_\text{Li} )!)$. 
Summing over $N_\text{Li}$ from $0$ to $16$ leads to 65,536 total configurations per system/phase. 
We do not consider mixtures of phases (domains) in a monolayer nor the possibility that two monolayers could belong to a different phase.

For each of the three MoX$_2$ systems and for both H and T' phases we randomly sample the configurational space such that the configurations with $N_\text{Li}$ intercalated Li spanning the range from $0$ to $N_{sites}= 16$ are roughly equally represented.
Each configuration is optimized by minimizing the stress tensor and relaxing the atomic forces to less than $0.005$ eV/ \AA.
During optimization the Brillouin zone is sampled with a 4$\times$4$\times$1 k-point grid.
Convergence in the electronic density of states of the relaxed structures is achieved with a 10$\times$10$\times$1 k-point grid. We do not include spin-orbit coupling due to the associated computational cost and because it is not expected to affect significantly the structural relaxation; we checked that spin-orbit coupling has an insignificant impact on the density of states near the Fermi level.

\subsubsection{Surrogate models}
Obtaining the total energy of all possible distinct initial configurations is computationally too expensive with {\it ab initio} DFT even for the relatively small 4$\times$2 supercell.
To this end we explore several surrogate models that can sample exhaustively the entire configurational space given DFT training data. 

The first method we consider is cluster expansion (CE) \cite{sanchez1984generalized,de1994cluster,li1994lattice}, a classical technique based on a linear coefficient model with a nearest neighbor expansion
\begin{equation}
E = E_\text{CE}(n_b; c_a)  =  c_0 +  \sum_a c_a  f_a(n_b)
\end{equation}
where the features $f_a$ are formed from a polynomial expansion of binned nearest neighbors $n_b$.
Here $n_b$ is the total number of neighbors in the $b$-th bin.
The bins we consider are annular rings with width small enough such that they contain one type of neighbor.
In a $n$-th order expansion $f_a$ are complete with powers of $n_a$ up to $n_{CE}$.
The coefficients $c_a$ are found via regression to training data. 
In the standard form of CE, the polynomial order is set to $n_{CE}=1$; we explored higher order polynomials as well.

We also used several machine techniques such as a basic multilayer perceptron (MLP) neural network (NN) as well as the more complex crystal graph convolutional neural network (CGCNN) \cite{xie2018crystal} and image-based transfer learning with a convolutional neural network \cite{Matlab}.
The MLP model is a feed-forward model of densely connected layers.
The output of each preceding layer (starting with the selected inputs) is transformed by a linear weight matrix and then a non-linear activation function.
The coefficients of the linear transformation of each layer comprise the parameters of the model.
We trained the parameters of the MLP using the neighbor counts $n_b$ as inputs and the corresponding energies as outputs.
The CGCNN formalism is more complex and is based on a universal and interpretable representation of crystalline materials.
An input crystallographic structure $\chi_i$ is encoded as a graph whose nodes, corresponding to atoms, are connected by edges, corresponding to bonds, with surrounding atoms.
The CGCNN architecture then applies a nonlinear graph convolution function that updates the atoms' feature vectors with information from the surrounding atoms and bonds.
A pooling layer then produces an overall feature vector for the entire crystal that satisfies invariance with respect to atom indexing (permutational invariance) and number of atoms in the unit cell (size invariance).
Similar to the MLP training, we optimized the parameters of the CGCNN to fit the DFT predicted energy.
We also applied a transfer learning technique using a pre-trained convolutional neural network (CNN), {\it googlenet} \cite{szegedy2015going}, originally used for image recognition.
We replaced the classification last layer of this 22-layer deep CNN with a new regression layer and re-trained the network on predicting energies based on input images of the non-self-consistent electrostatic potential projected on a plane that passed through the unrelaxed Li-ion layer.
The non-self-consistent electrostatic potential was selected since it is inexpensive to compute and the projection was necessary since {\it googlenet} operates on two-dimensional images.

\begin{figure}[h!]
\centering
\subfloat[total energies]
{\includegraphics[width=0.5\textwidth]{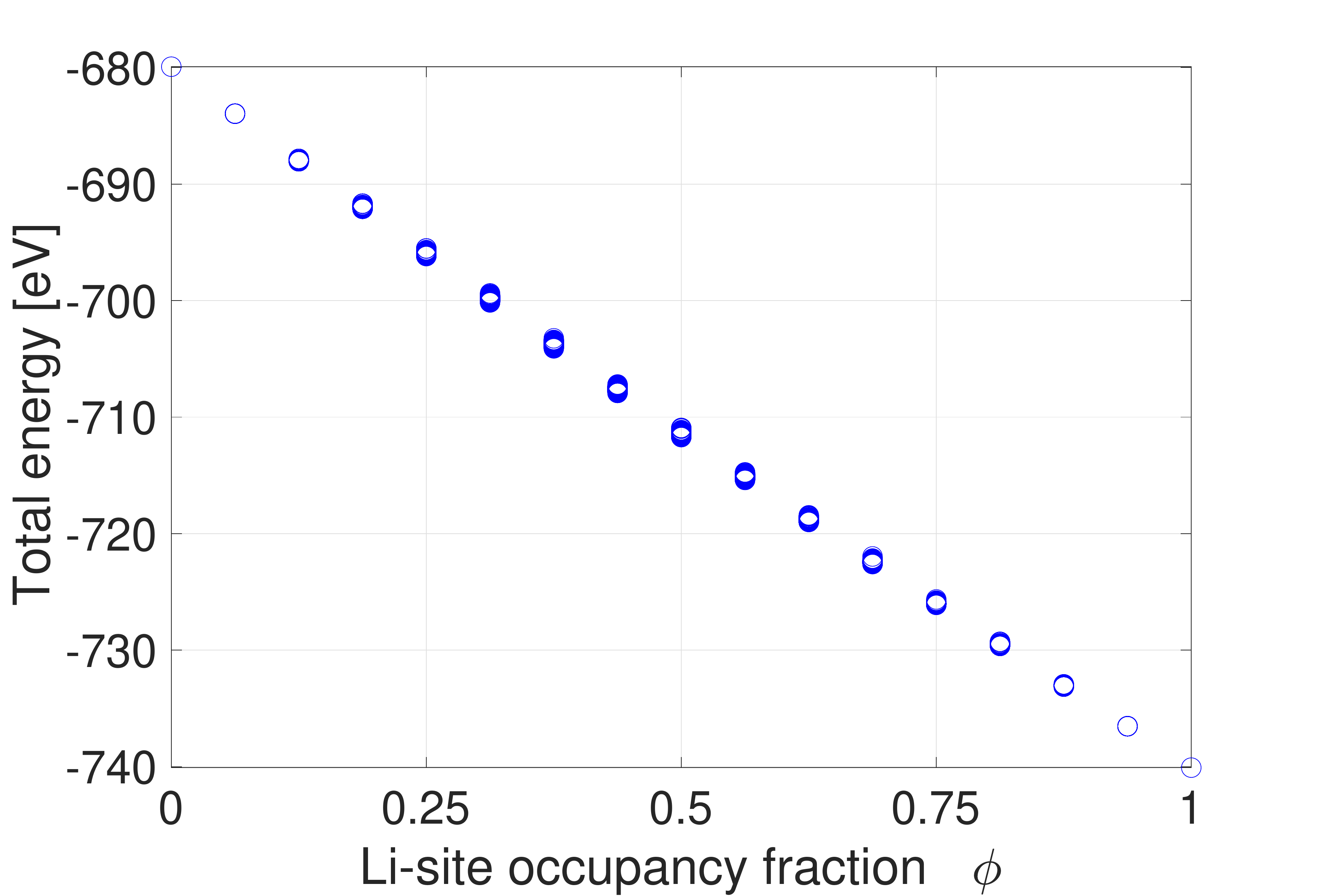}}
\subfloat[formation energies]
{\includegraphics[width=0.5\textwidth]{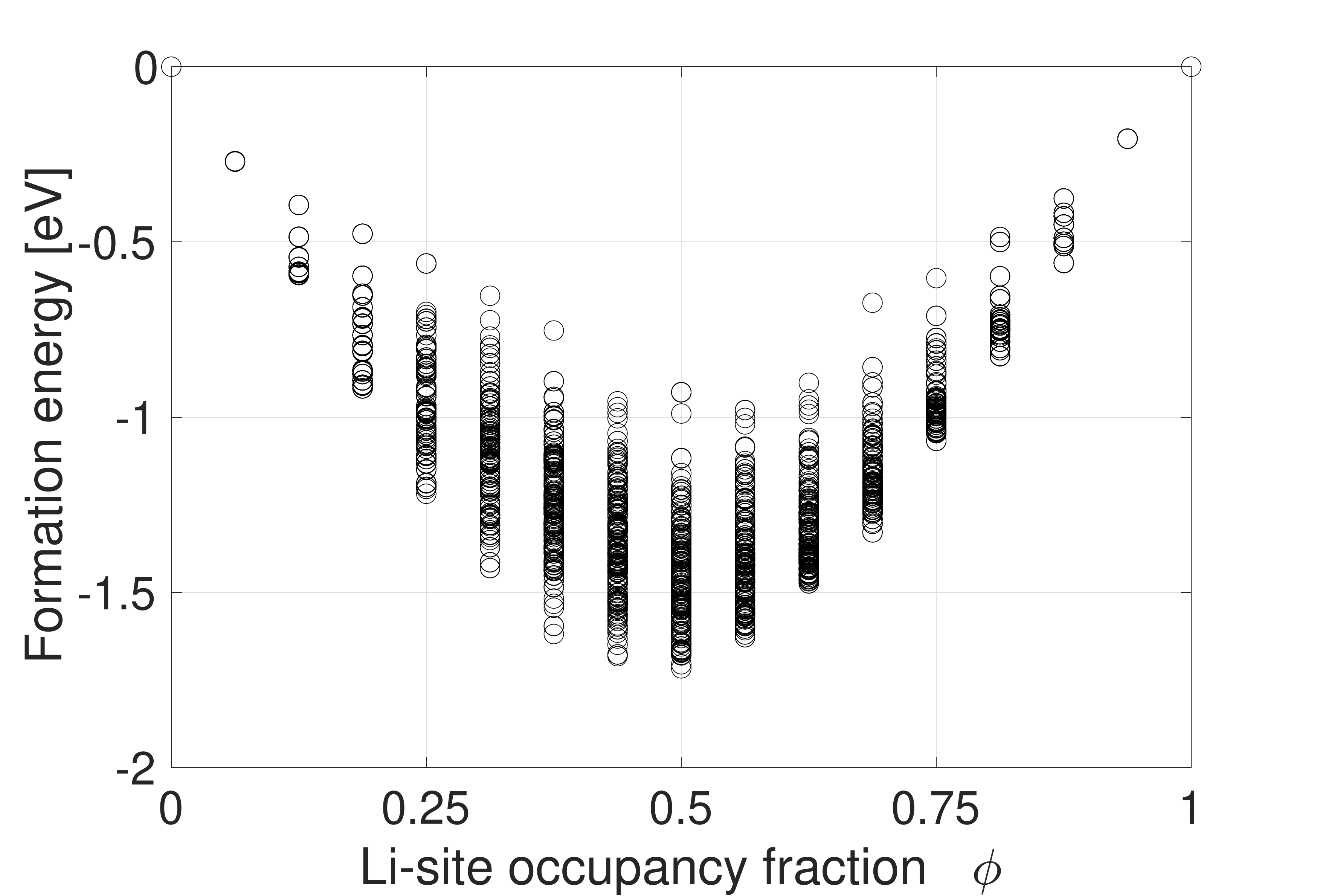}}
\caption{DFT (a) total energies and (b) formation energies of bilayer MoS$_2$ T' phase as function of the  number of intercalated Li ions per supercell.}
\label{fig:form_E}
\end{figure}

For all these approaches the training set consisted of hundreds to thousands of input-output pairs. The inputs correspond to unrelaxed Li-intercalated MoX$_2$ structures.
Prior to the training models of the energy as a function of configuration,  we transformed total energies $E$ obtained from DFT into formation energies $\tilde{E}$ according to a linear transformation \eref{eq:formE}.
As seen in \fref{fig:form_E}, this transformation reduces the energy range of the outputs by more than an order of magnitude, from about $50$ eV (the total energy decreases with increasing Li-site occupancy fraction because the Li intercalation energy is positive  \cite{lu2020lithium}) to about $3$ eV, facilitating better convergence during the training process.

\section{Results: phase stability and electronic structure evolution}

With the methods described in the previous section we are able to calculate the free energy difference between the H and T' phases of MoS$_2$, MoSe$_2$ and MoTe$_2$ and estimate the phase transition as a function of Li site occupancy.
Furthermore, we provide the DOS at the Fermi level to indicate where changes in conductivity  occur with increasing Li occupancy.

\subsection{Free energies and phase diagrams}

Depending on the experimental conditions, the phase stability of Li ion intercalated bilayer TMDs should be computed with the closed system, NT, ensemble or the open system, $\mu$T, ensemble.
To this end, the calculated energies were used to construct the NT and $\mu$T potentials as well as the statistical average of the Li site occupancy $\langle \phi \rangle$ as function of Li chemical potential. 
Analysis of the errors induced by using a 4$\times$2 system are detailed in Appendix A.
This analysis indicates the finite size errors are negligible.
Appendix B describes a comparison between the surrogate models in representing formation energies and DOS.
Given the task of representing formation energies as a function of Li site occupancy, we show that classical cluster expansion (CE) is sufficient and competitive with the modern machine learning surrogate models we implemented in both accuracy and computational expense.
The lower complexity CE model does well since the main input is the on-site occupancy of a small lattice and results in mean absolute errors (MAEs) of $\approx$ 50 meV.
Compared to the MLP model with same features the main difference is a global polynomial basis for the CE model versus the piecewise basis of a typical NN.
More sophisticated representations such as CGCNN or an image-based NN would differentiate themselves if the configurational space were more complex, for instance if there was significant rearrangement of the bilayer with relaxation.

\begin{figure}[h!]
\centering
\subfloat[MoS$_2$]
{\includegraphics[width=0.6\textwidth]{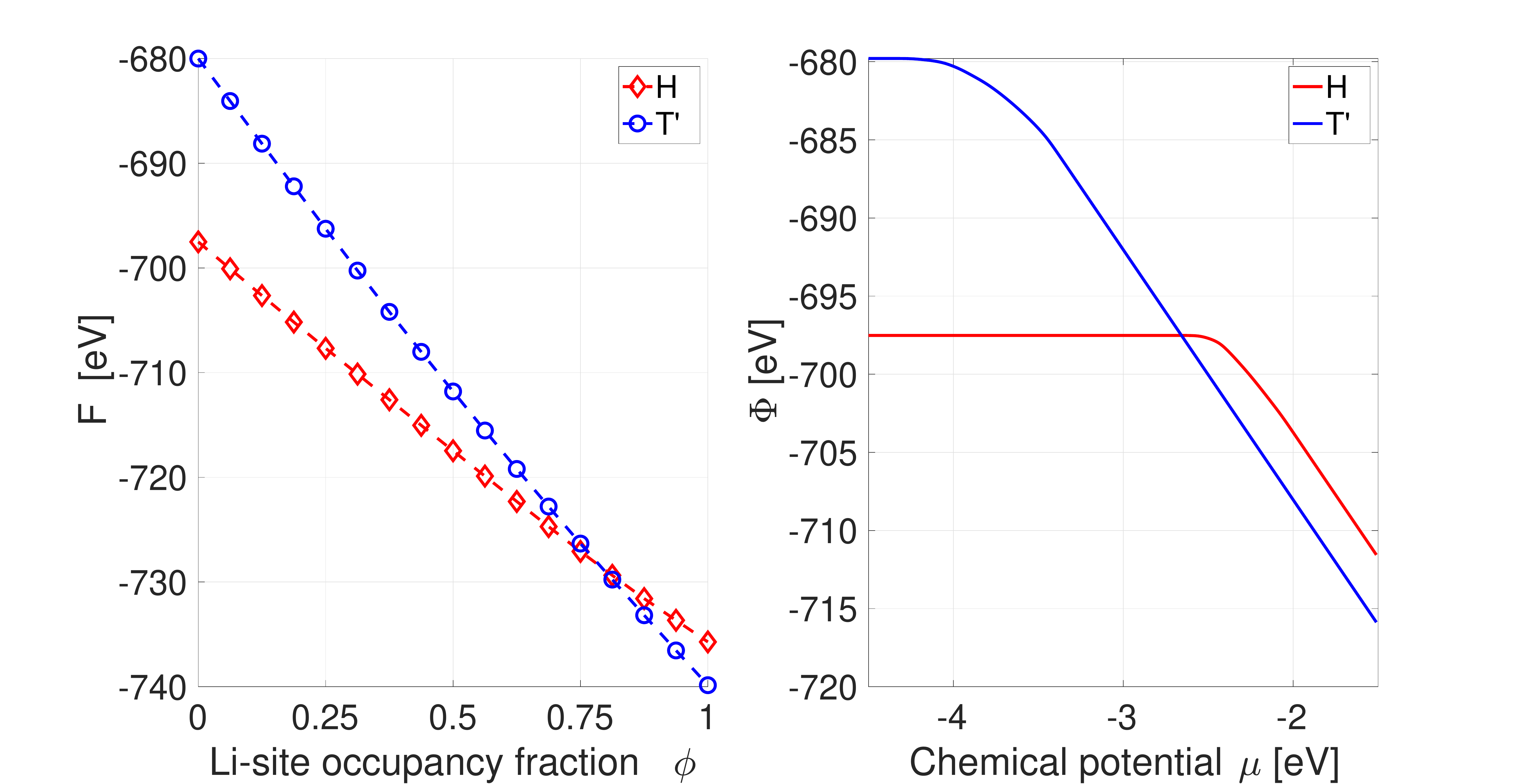}}\\
\subfloat[MoSe$_2$]
{\includegraphics[width=0.6\textwidth]{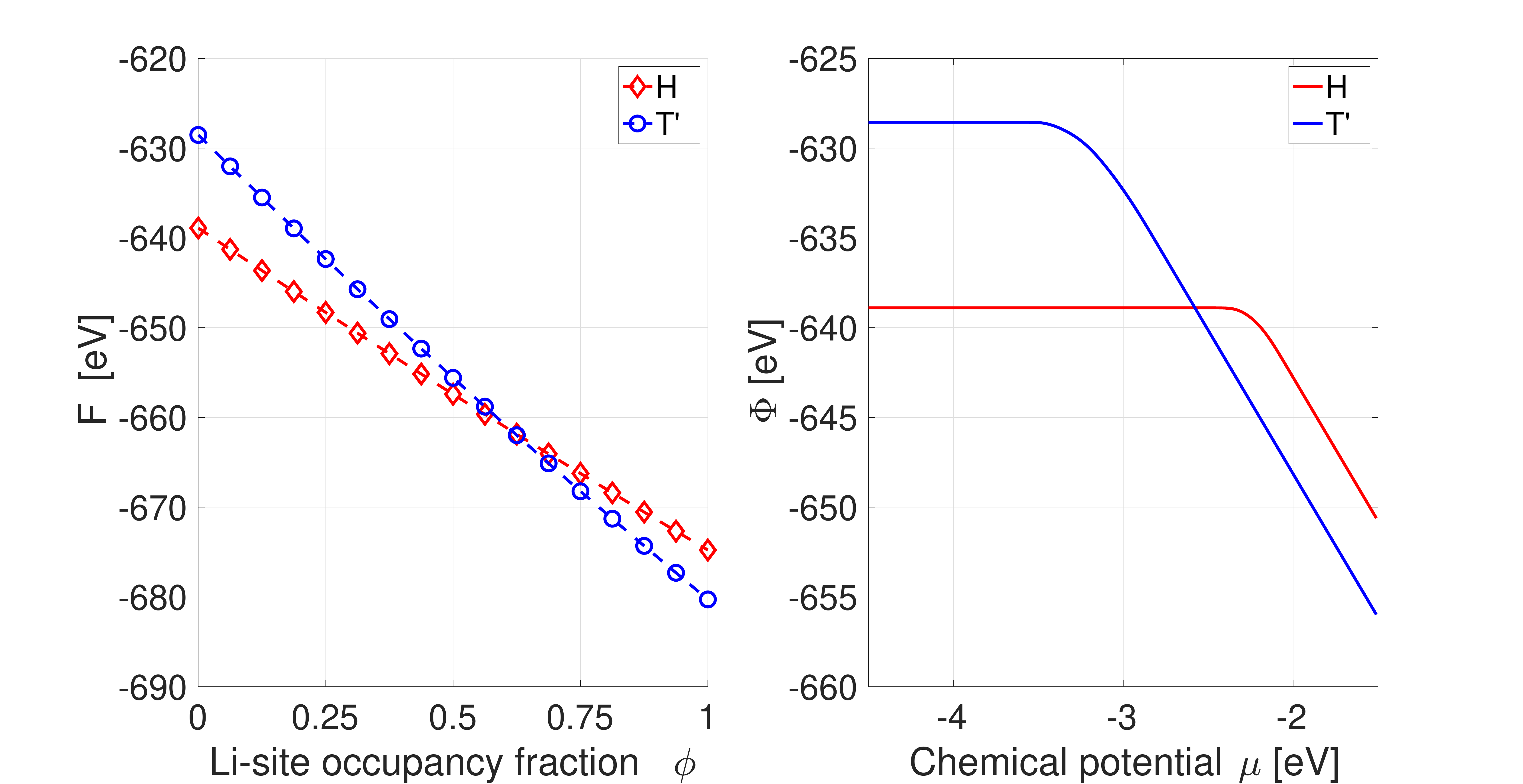}}\\
\subfloat[MoTe$_2$]
{\includegraphics[width=0.6\textwidth]{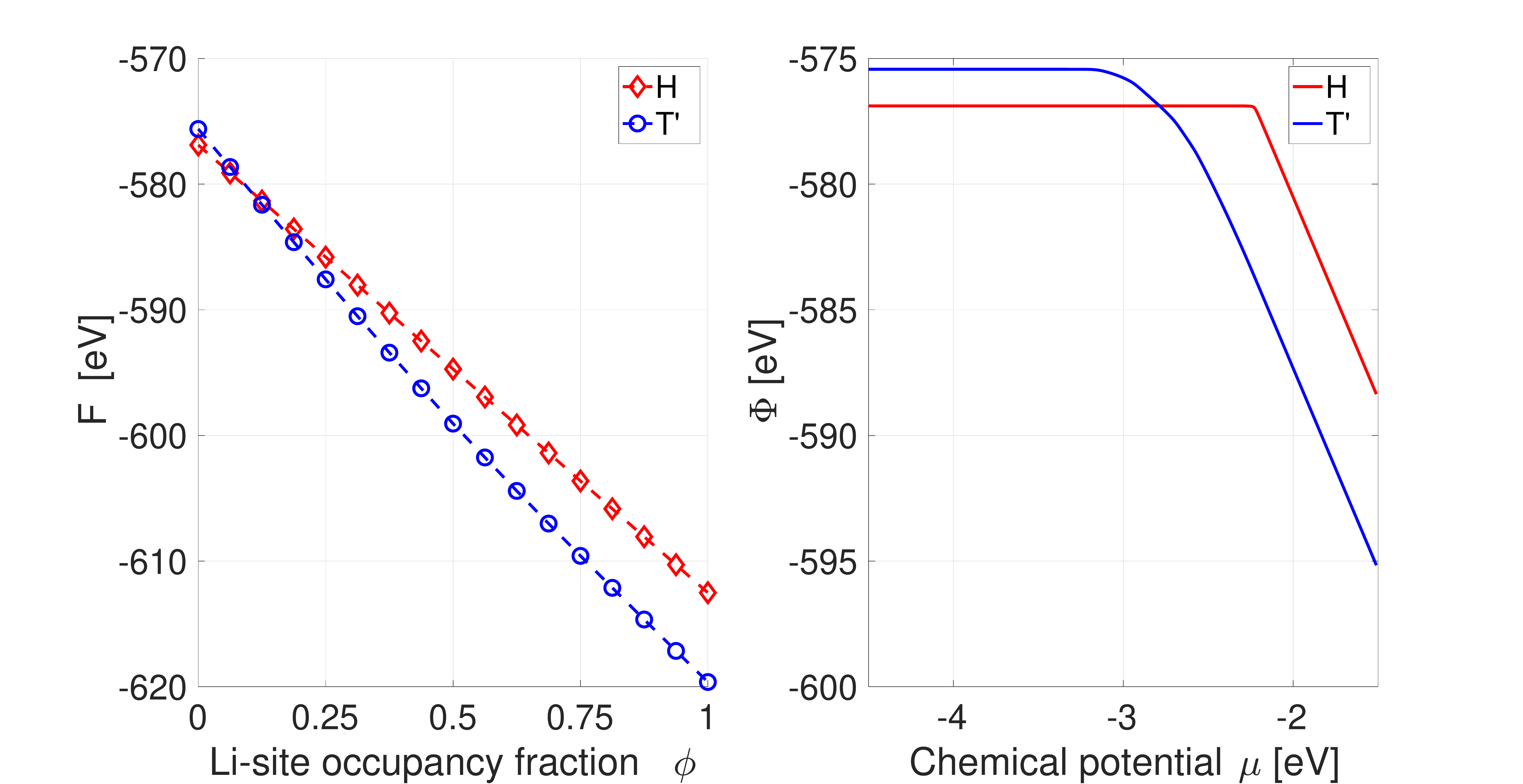}}
\caption{NT (left panels) and $\mu$T (right panels) free energies for: (a) MoS$_2$,  (b) MoSe$_2$, and  (c) MoTe$_2$ Li ion intercalated bilayers. $T$=300~K.
 }
\label{fig:Free_energies}
\end{figure}

\fref{fig:Free_energies} shows the NT free energy and $\mu$T free energies for three Li ion intercalated bilayer systems:  MoS$_2$, MoSe$_2$, and MoTe$_2$, at room temperature $T$=300~K.
It appears that the T' phase is preferred at high ion concentrations/low chemical potentials, as its free energy becomes lower than that of the H phase.
Interestingly, as the chalcogen atom X gets heavier, the concentration range of stable T' phases becomes larger and the threshold ion concentration where the H-T' transition takes place gets smaller.

\clearpage

\begin{figure}[h!]
\centering
\subfloat[NT ensemble]
{\includegraphics[width=0.45\textwidth]{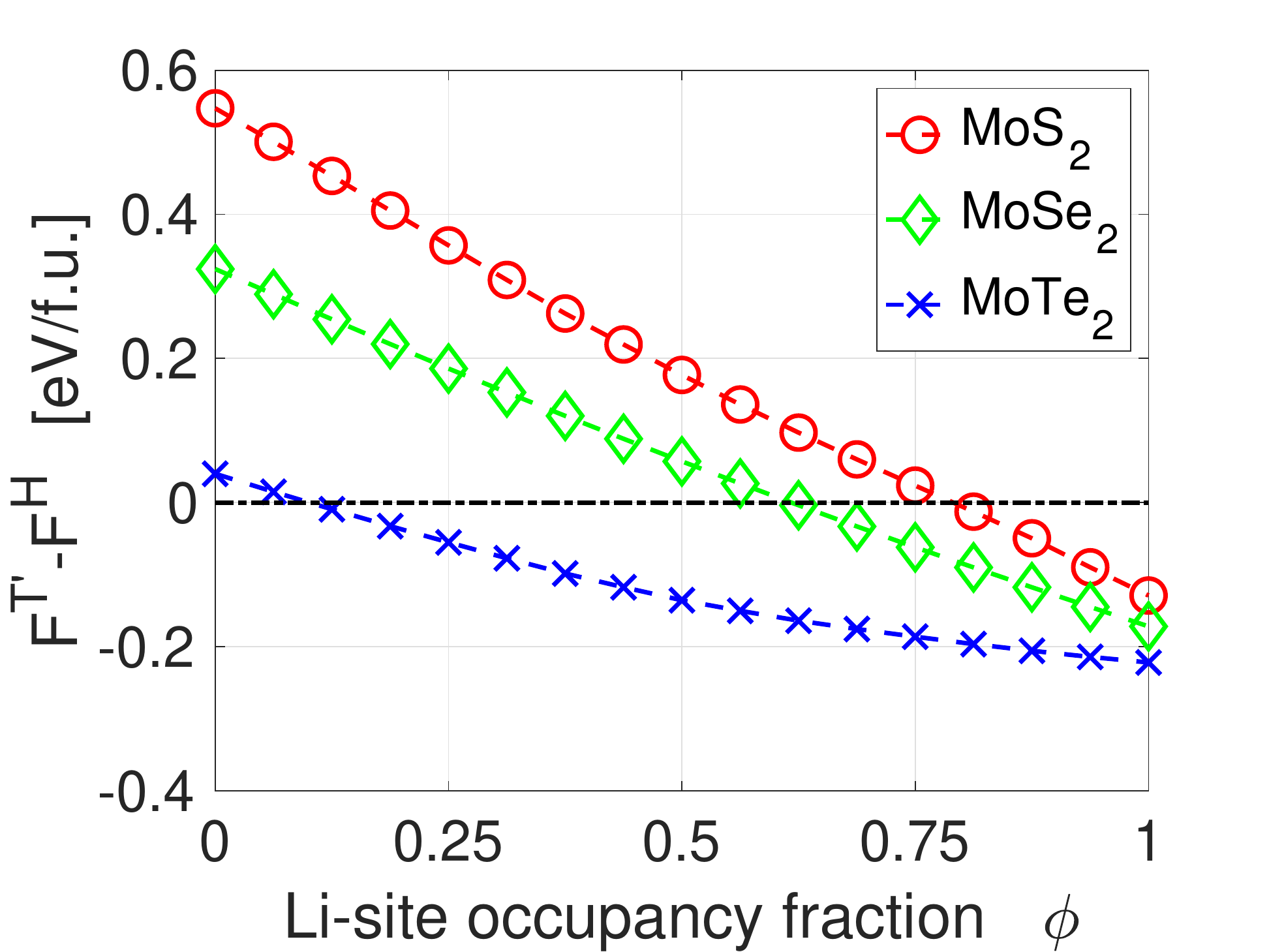}}
\subfloat[$\mu$T ensemble]
{\includegraphics[width=0.45\textwidth]{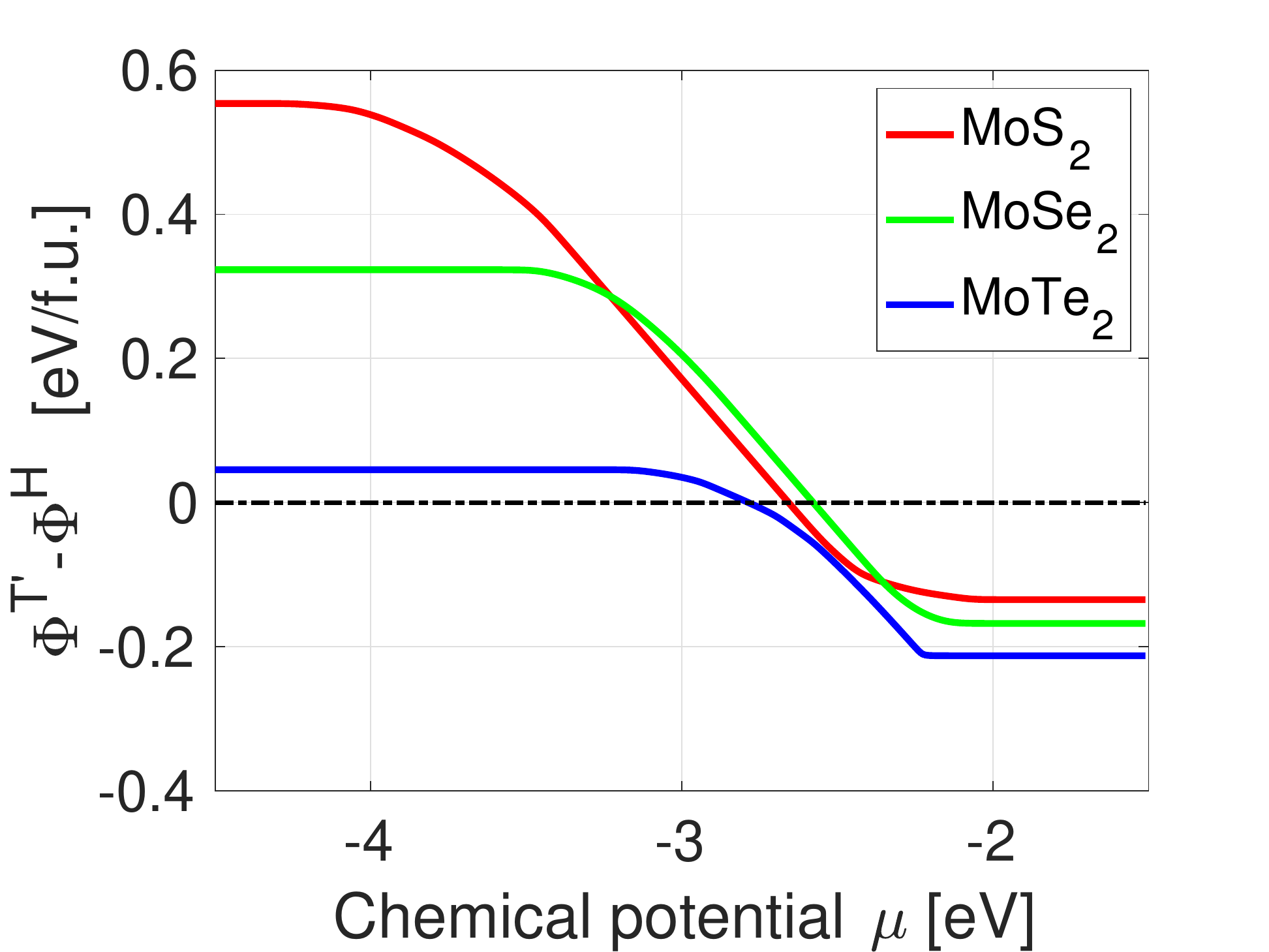}}
\caption{Free energy differences between competing phases for three Li ion intercalated bilayer MoX$_2$ TMDs (X = S, Se and Te) obtained within the: (a) NT,  (b) $\mu$T ensemble at  $T$=300~K.
}
\label{fig:Phase_diagrams}
\end{figure}

This trend can be clearly seen in \fref{fig:Phase_diagrams} which shows the phase diagram at $T$=300~K for the three bilayer systems.
\fref{fig:Phase_diagrams}a shows that the stabilization energy (the free energy difference between T' and H phases at $\phi=0$) decreases from $\approx0.55$ eV/formula unit for MoS$_2$, to $\approx 0.35$ eV/formula unit for MoSe$_2$ and finally decreases to only $\approx 0.05$ eV/formula unit in the case of MoTe$_2$.
A similar trend has been discovered in the case of Li ion adsorption onto monolayer MoX$_2$ \cite{li2016ferroelasticity}.
Besides the stabilization energy, the threshold Li concentration where the H$\leftrightarrow$T' transition happens decreases as well, from $x\approx 0.4$ ions/formula unit for MoS$_2$, to $x\approx 0.3$ ions/formula unit  for MoSe$_2$ and finally to $x\approx 0.05$ ions/formula unit in the case of MoTe$_2$.
(Recall $x = 1/2 \phi$.)
The total change in free energy from zero occupancy to site saturation decreases with the mass of X, from $\approx0.7$ eV/formula unit for MoS$_2$, to $\approx$0.5 eV/formula unit for MoSe$_2$, to $\approx$0.25 eV/formula unit for MoTe$_2$, indicating the relative stability of the H and T' phases decreases over this range.
Similarly, \fref{fig:Phase_diagrams}b shows that the Li chemical potential threshold is lowest when the chalcogen atom X is the heaviest among the three cases considered, \ie X=Te. It also appears that the threshold $\mu$ is higher for X=Se than for X=S case, however this particular ordering may depend on the accuracy with which one evaluates the free energy within the $\mu T$ ensemble.

The results shown in \fref{fig:Free_energies} and \fref{fig:Phase_diagrams} are obtained within CE but the improvement with respect to direct use of the DFT data in \eref{eq:zNT} and \eref{eq:zmuT} is marginal \ie estimating the energies of all the states improves the smoothness of the ramp-like shape of the free energy profiles and possibly slightly alters the phase crossings.
We find that differences between the surrogate methods and DFT are most evident in the evaluation of the average Li-site occupancy fraction $\langle \phi \rangle$  within the $\mu T$ ensemble, as shown next.
 
\begin{figure}[h!]
\centering
{\includegraphics[width=0.5\textwidth]{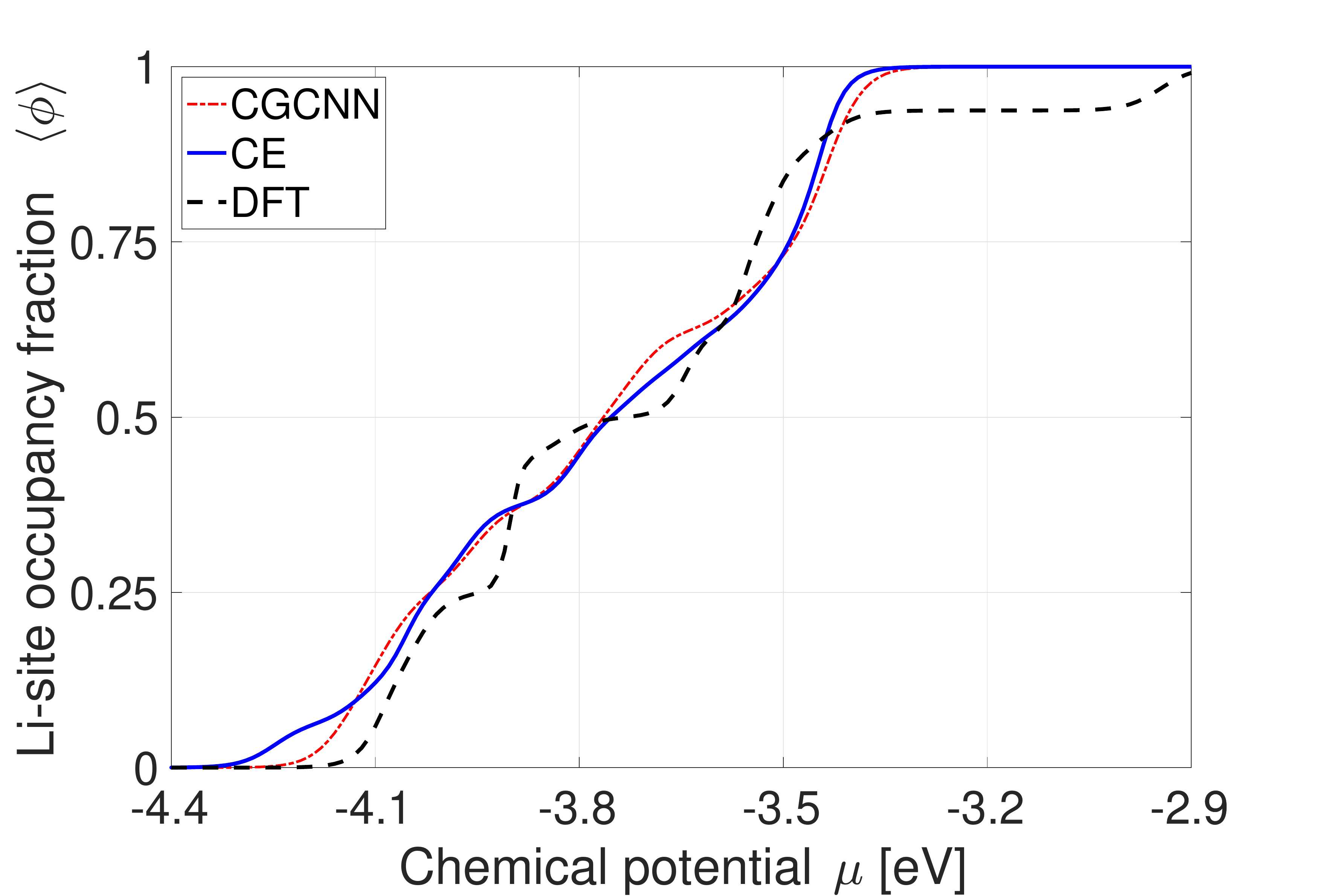}}
\caption{$\mu$T statistical average of the Li-site occupancy fraction  $\langle\phi\rangle$ as function of Li chemical potential for bilayer MoS$_2$ T' phase. $T$=300~K. 
}
\label{fig:avgN_mu}
\end{figure}

\fref{fig:avgN_mu} shows  $\langle \phi \rangle$ \vs the Li chemical potential $\mu$ for a relatively large range of  $\mu$, calculated based on energy predictions made via direct use of the DFT data, or the CE and CGCNN models using 65,536 configurations of the bilayer MoS$_2$ T' phase.
We estimated the chemical potential in a Li-rich environment $\mu^{Li-rich}$ from the DFT binding energy of elemental Li metal and find that $\mu^{Li-rich}\approx -2$ eV. 
Both CE and CGCNN predict that in a Li-rich environment intercalation is complete, \ie the Li ion concentration saturates to $\langle \phi \rangle =1$.
The two models agree overall, which correlates with the fact  that they also yield a similar $\textrm{MAE}\approx 50$ meV in predicted energies. 
It is difficult to draw conclusions regarding which model is most accurate in predicting $\langle \phi \rangle$ versus $\mu$ and this distinction may be not important as 50 meV is considered the consistency level between different DFT functionals.

\subsection{Density of electronic states}

\subsubsection{Trends in the DOS with increasing ion concentration}

To  make predictions regarding the electronic conductivity  of Li ion-intercalated bilayer TMDs as function of ion concentration we consider the electronic density of states (DOS) near the Fermi level (E$_F$) for each system/phase. 
Indeed, within band theory, DOS(E$_F$) is directly proportional to the low voltage bias electronic conductivity. 

\begin{figure}[h!]
\centering
\subfloat[MoS$_2$]
{\includegraphics[width=0.45\textwidth]{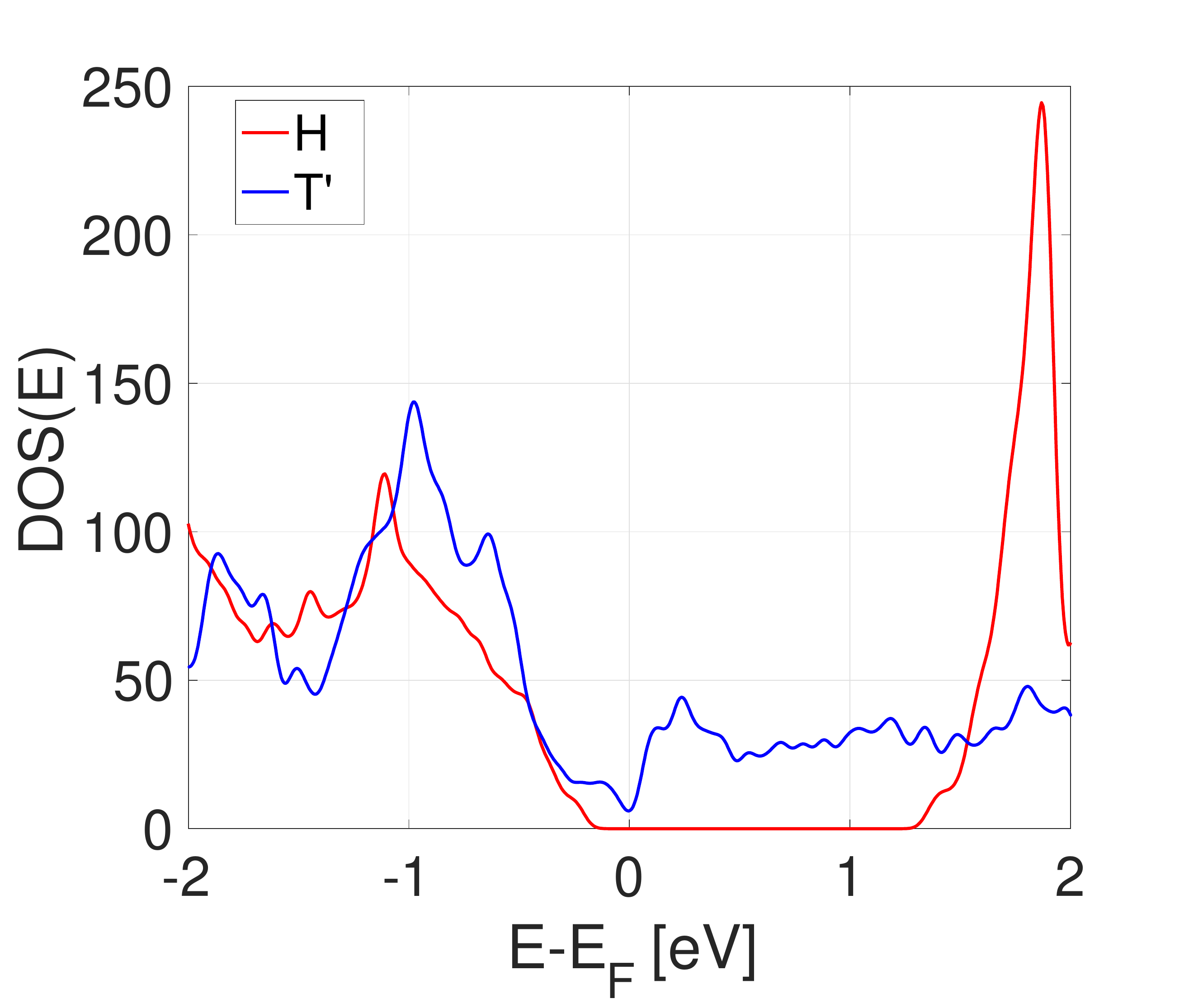}}
\subfloat[MoTe$_2$]
{\includegraphics[width=0.45\textwidth]{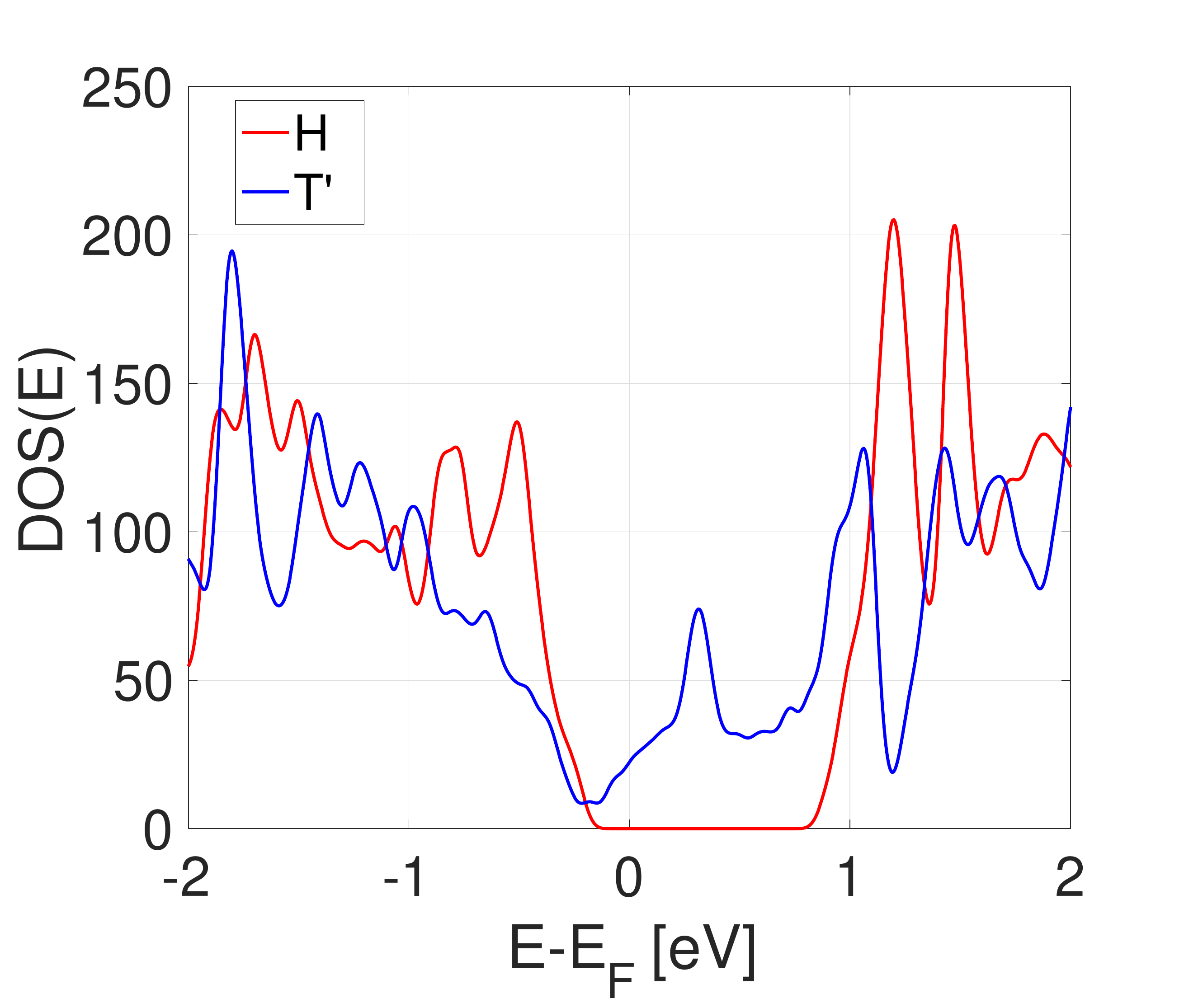}}
\caption{Electronic density of states as function of the electron energy measured from the Fermi level E$_F$ for (a) bilayer MoS$_2$ and (b) bilayer MoTe$_2$. Systems with no Li ion intercalation.}
\label{fig:dosE_MoS2Te2}
\end{figure}

\begin{figure}[h!]
\centering
\subfloat[$N_\text{Li}$=0]
{\includegraphics[width=0.45\textwidth]{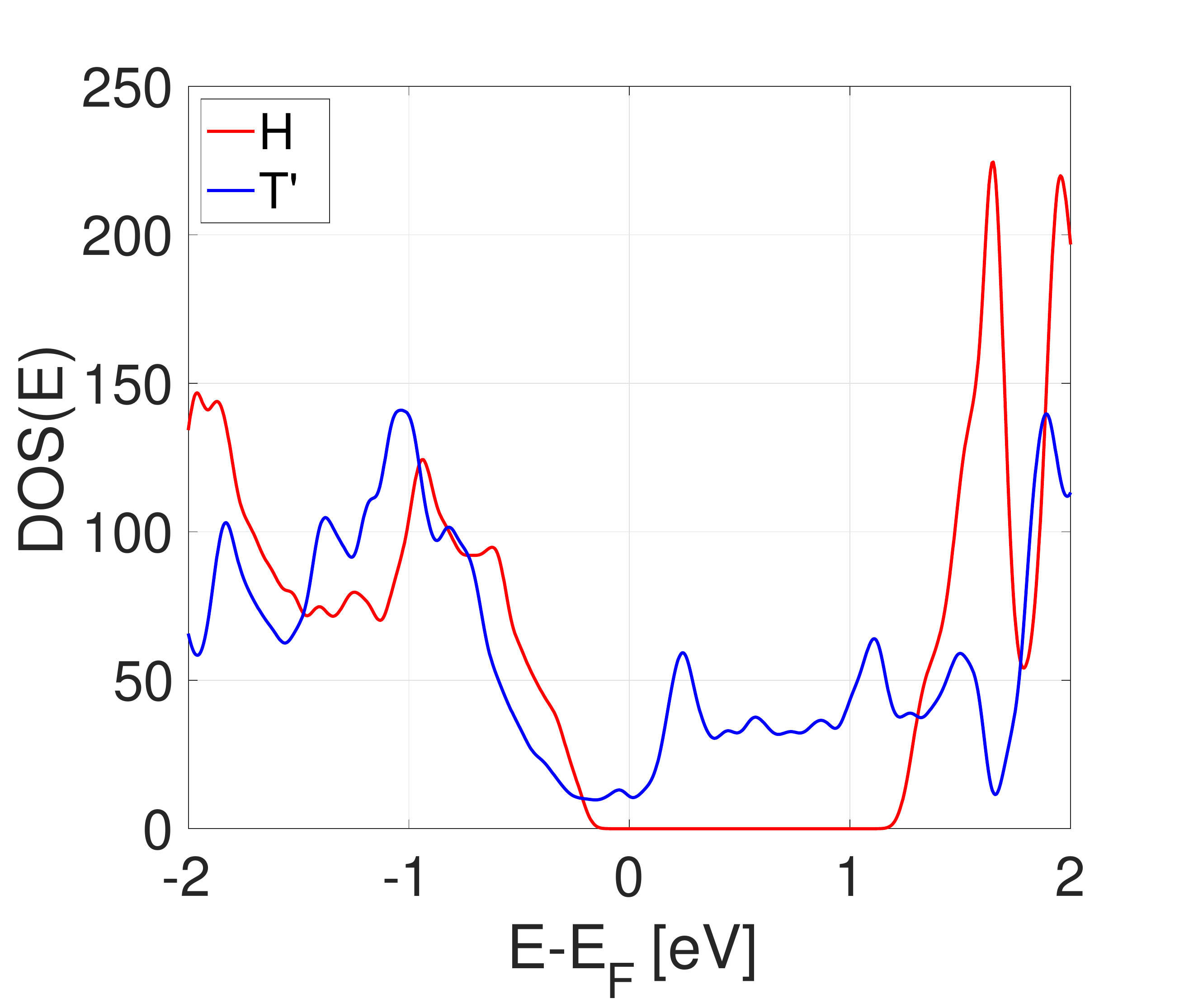}}
\subfloat[$N_\text{Li}$=1]
{\includegraphics[width=0.45\textwidth]{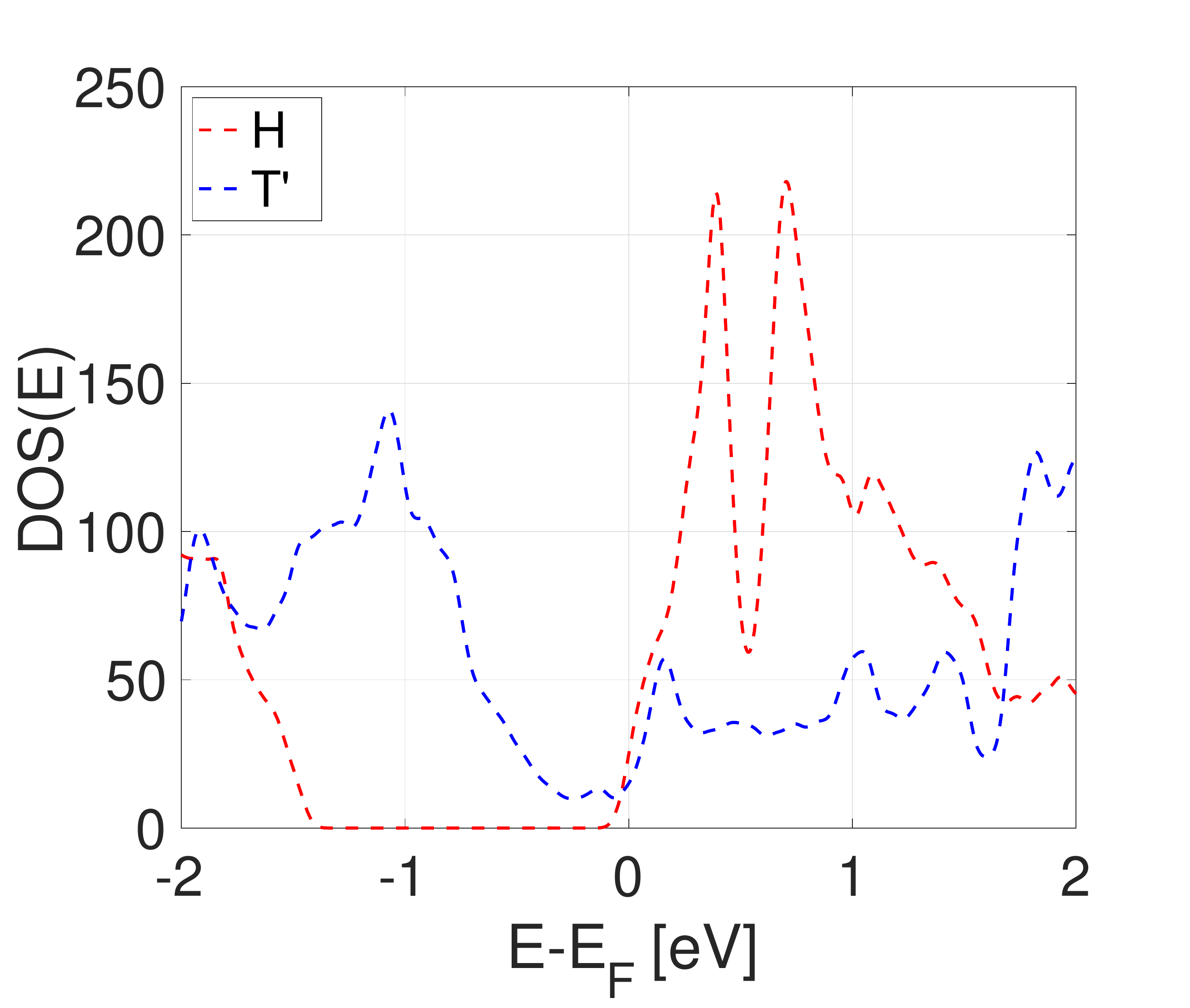}}
\caption{Electronic density of states as function of the electron energy measured from the Fermi level E$_F$ for bilayer MoSe$_2$ with: (a) no Li ion intercalation, (b) one intercalated Li ion per supercell.}
\label{fig:dosE_MoSe2}
\end{figure}

\fref{fig:dosE_MoS2Te2} and \fref{fig:dosE_MoSe2}(a) show the DOS for each system/phase in the absence of Li ion intercalation.
We observe that the H is semiconducting while the T' phase is semimetallic.
In the H case the band gap decreases with increasing mass of the chalcogen atom X.
  
To see the impact of Li ion intercalation on the DOS we consider in \fref{fig:dosE_MoSe2}(b) the case of MoSe$_2$ with one Li ion/supercell intercalated in the bilayer.
We note that for this doping level the DOS of both phases do not renormalize appreciably.
Instead, they can be obtained from the no-intercalation case by a simple shift-up of the Fermi level.
Due to the semiconducting band gap in the H case, the shift to the left of the DOS is correspondingly large $> 1$ eV.
In the semi-metallic T' case the DOS shift is much smaller.
Thus, the change in the DOS(E$_F$) with low Li ion concentration is expected to be small in the T' phase case and large for the H phase.

\begin{figure}[h!]
\centering
\subfloat[several $\phi$, fully relaxed]
{\includegraphics[width=0.45\textwidth]{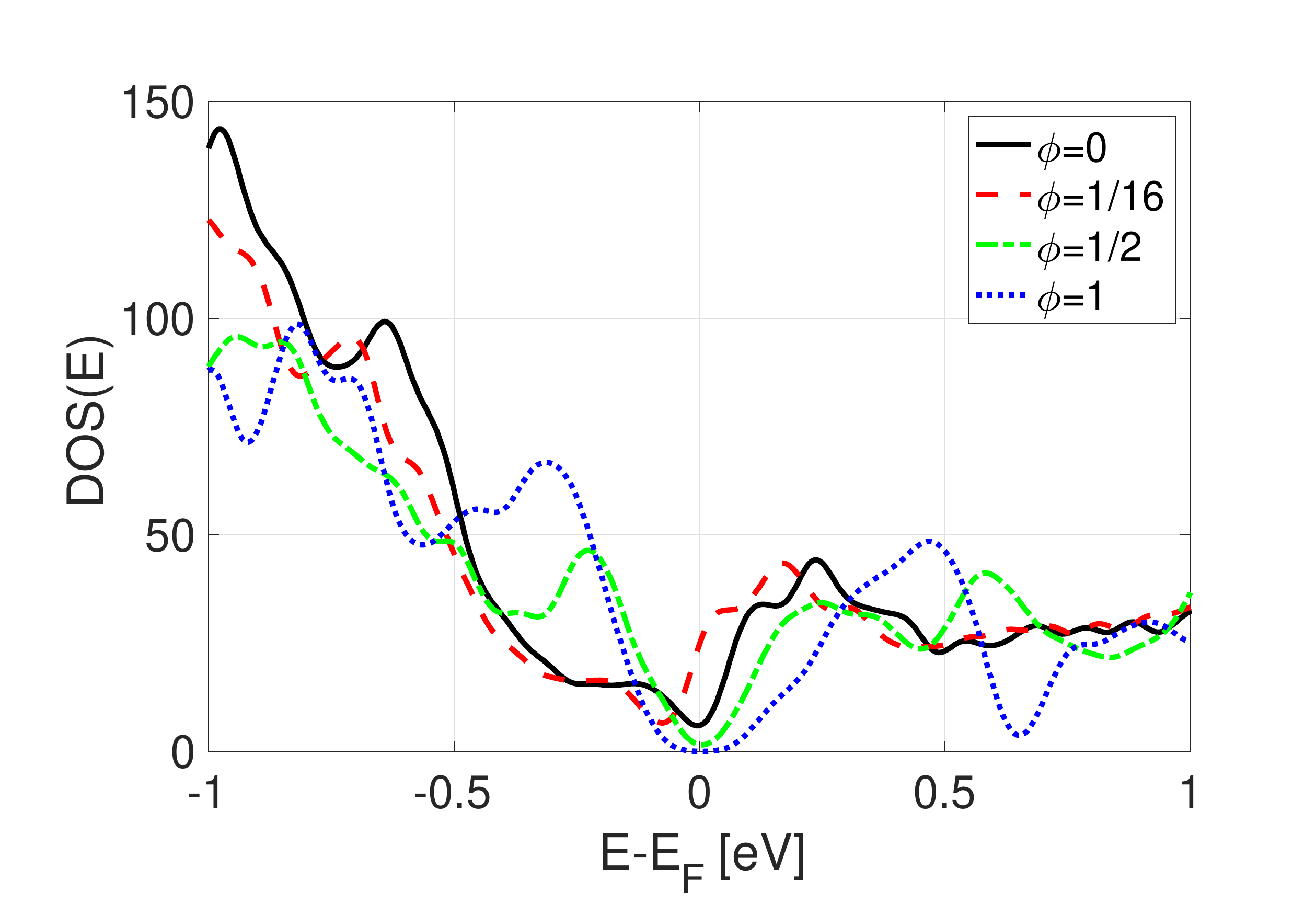}} 
\subfloat[$\phi=1$, full versus partial relaxation]
{\includegraphics[width=0.45\textwidth]{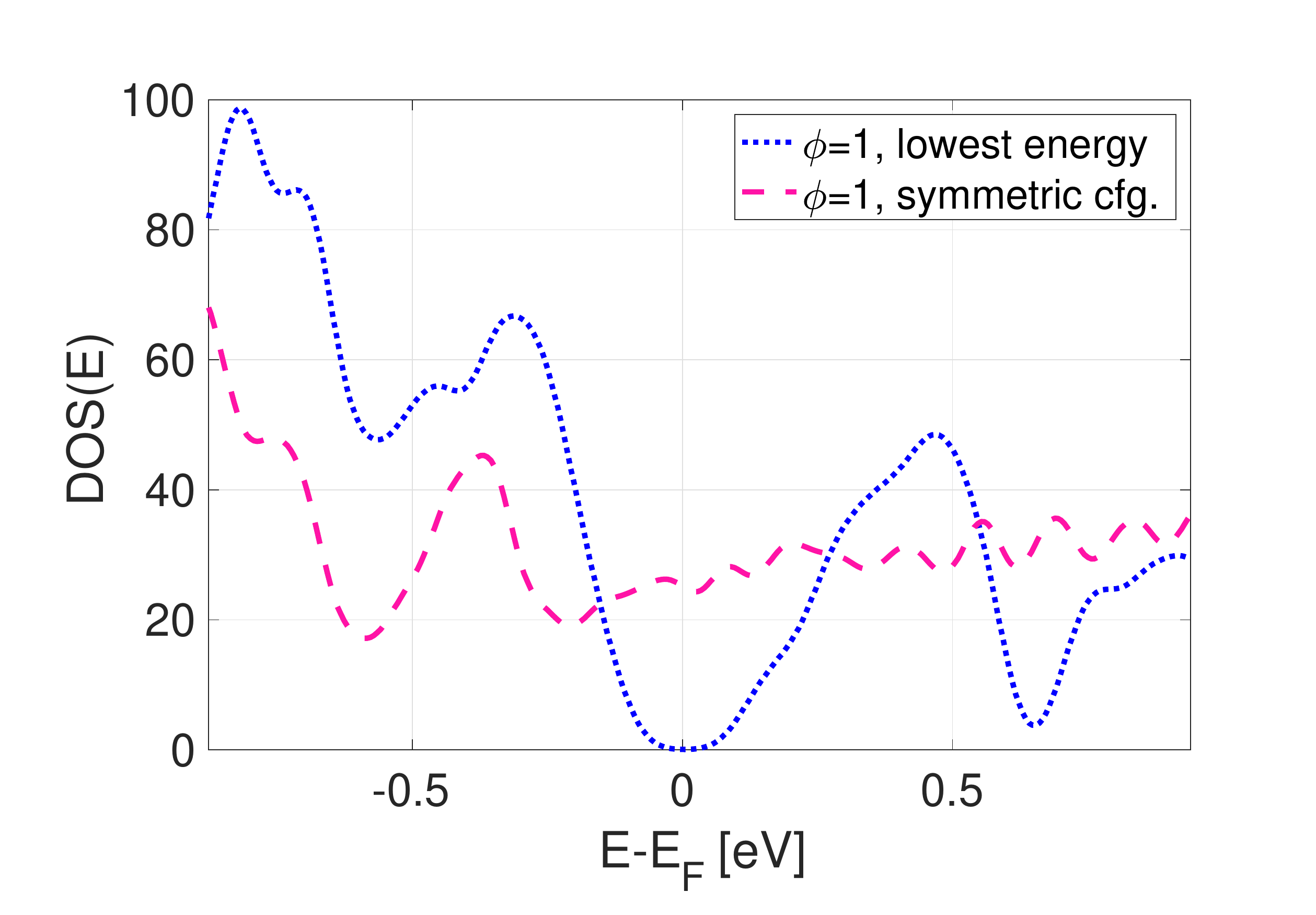}} 
\caption{Density of states of  Li ion intercalated bilayer MoS$_2$ T' phase for: (a) several Li-site occupancy fractions $\phi$ ranging from $0$ to $1$. (b)  $\phi=1$ (full Li intercalation) comparing the fully relaxed (lowest energy) case that leads to slight structural distortion versus the symmetric case where the $16$ Li ions are equivalent.
}
\label{fig:dosE_MoS2_Tp_n}
\end{figure}

With larger Li ion concentration it is possible that the DOS renormalizes, {\it i.e.} cannot be obtained from the no-intercalation ($N_\text{Li}=0$) case by a simple shift.
This can be seen in \fref{fig:dosE_MoS2_Tp_n}(a) which shows the DOS of bilayer MoS$_2$ (T' phase) for several intercalation levels.
As just mentioned, at low ion concentration the DOS can be obtained from the no-intercalation case by a simple, small shift (compare the black $\phi=0$ and red $\phi=1/16$ lines).
However, in the case of half-occupancy of the Li sublattice ($\phi=1/2$, $x=1/4$), the DOS shows very low DOS(E$_F$), clearly a renormalization feature that is not evident in the absence of Li.
At full occupancy of the Li sublattice ($\phi=1$, $x$=1/2), the renormalization is so strong that it induces a small gap in the bandstructure about the Fermi level.
The strong renormalization of the electronic structure that takes place at higher ion concentration for all three systems (especially for the T' phase) is not due to a simple electronic doping effect, but rather by geometrical relaxation/disordering of the bilayer MoX$_2$ atoms induced by the intercalated Li ions.
The disordering effect takes place even at full occupancy of the Li sublattice, where prior to atomic optimization the structure has translational symmetry.
This can be seen in \fref{fig:dosE_MoS2_Tp_n}(b) for the case of MoS$_2$ T' phase at full Li intercalation ($\phi=1$) showing that an opening of the band gap at the Fermi level occurs only upon full atomic relaxation that breaks the translational symmetry and renders the Li ions inequivalent. This disordering effect is subtle, with the mean distortion of atomic positions of $<0.015$ \AA. We note that periodic lattice distortions in TMDs have also been connected to a charge-density-wave formation mechanism \cite{doping2017,cdw1974,doping2017}.

\subsubsection{Evolution of DOS(E$_F$) with ion concentration}
The evolution of DOS(E$_F$) with ion concentration is obtained within the NT ensemble by statistically averaging the DFT calculated DOS(E$_F$) for each of the $\approx 300$ configurations per bilayer system/phase.
Here we do not use a surrogate model to estimate DOS(E$_F$) in the entire configurational space, but in Appendix B we show how two surrogate models (CE and NN) compare with DFT when representing the DOS.

\begin{figure}[h!]
\centering
\subfloat[MoS$_2$, H phase]
{\includegraphics[width=0.5\textwidth]{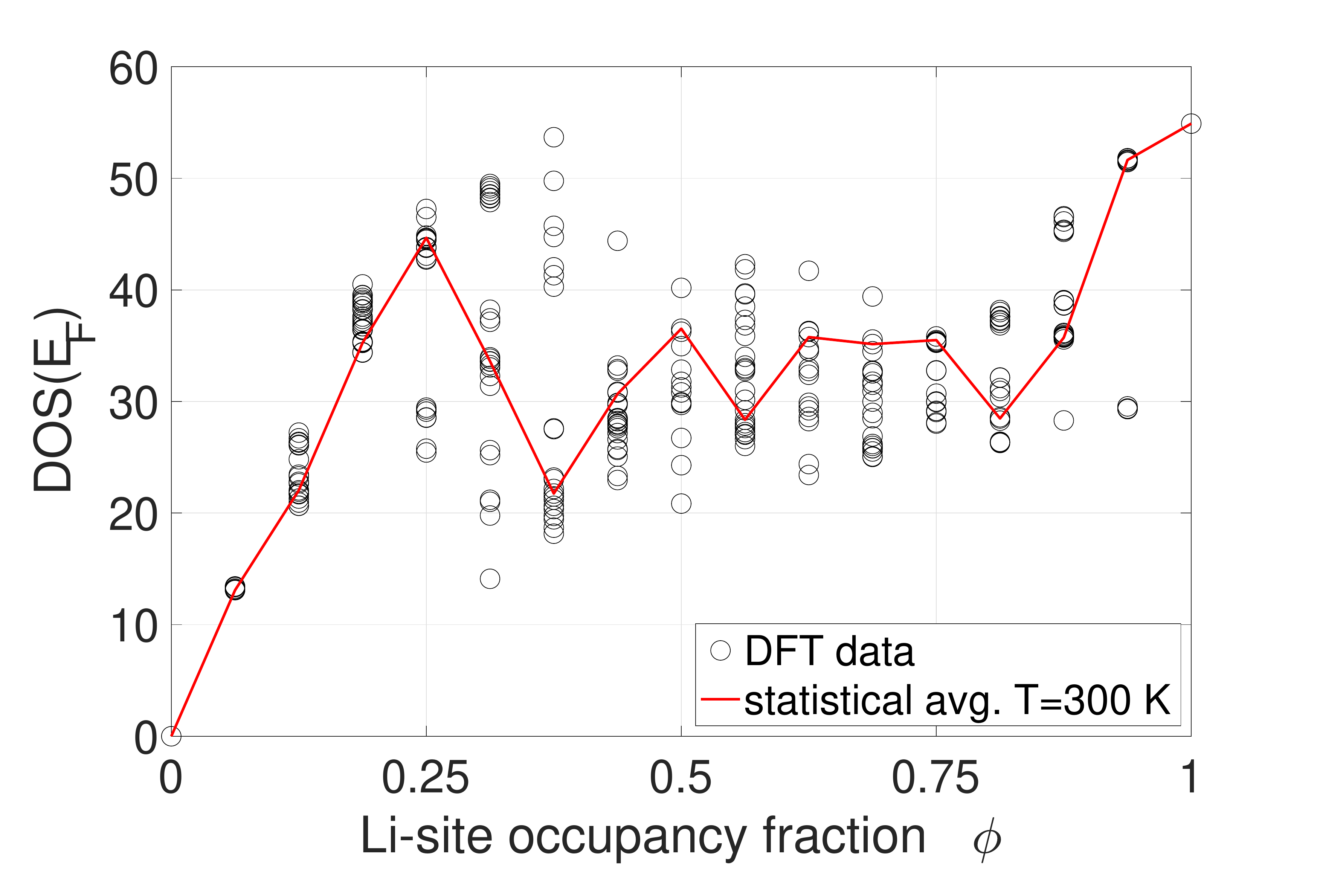}}
\subfloat[MoS$_2$, T' phase]
{\includegraphics[width=0.5\textwidth]{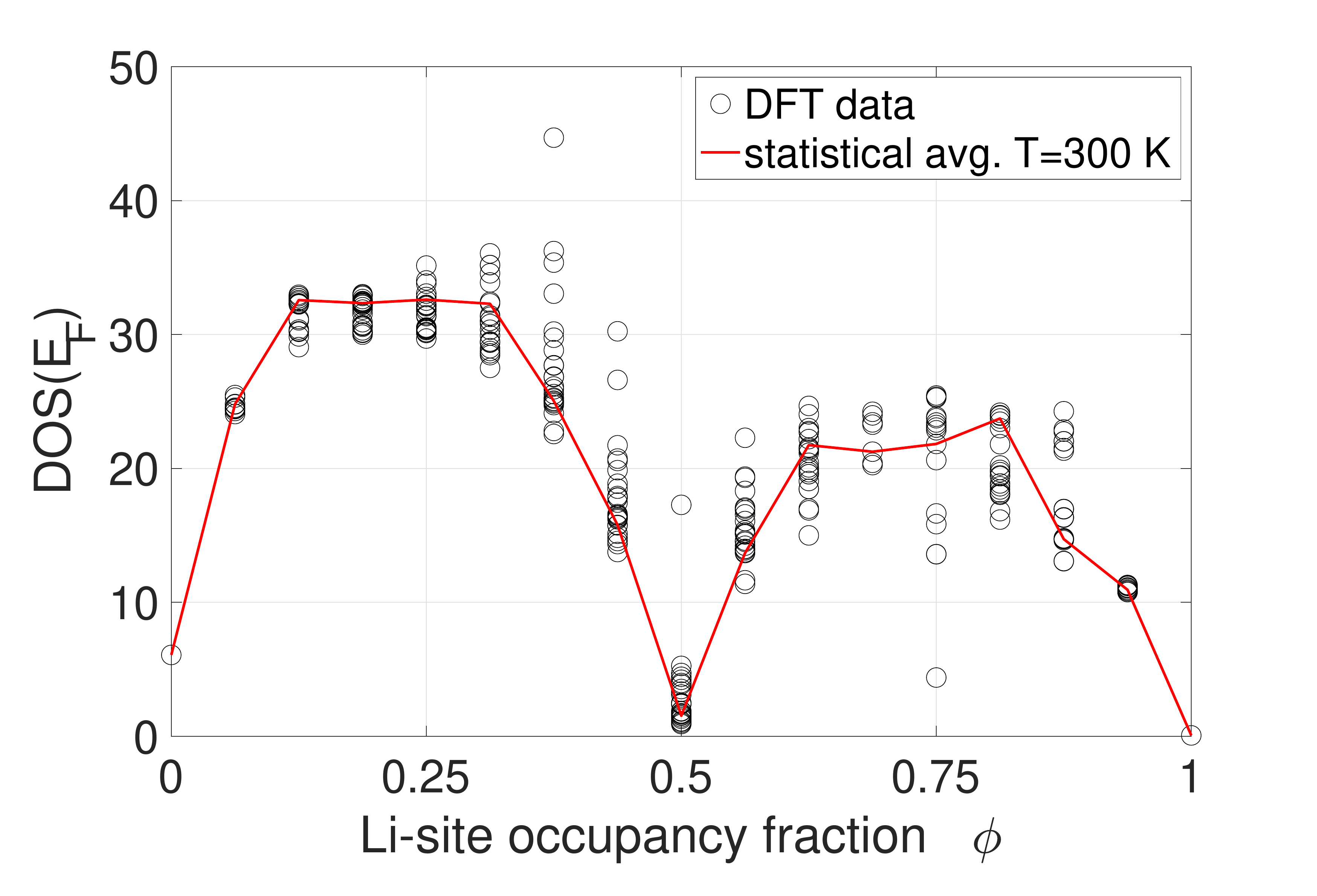}}
\caption{Density of states at the Fermi level of  Li ion intercalated bilayer MoS$_2$ for the two competing phases: (a) H,  (b) T' at $T$=300~K. 
}
\label{fig:dosE_MoS2}
\end{figure}

\fref{fig:dosE_MoS2} illustrates  our procedure for obtaining the statistically averaged DOS(E$_F$) as function of ion concentration from the raw DFT data for the case of the bilayer MoS$_2$.
We note the peculiar feature of very small DOS(E$_F$) for the T' phase at half Li sublattice occupancy.
It is known that one-dimensional metals are unstable against a distortion of the crystal lattice that lowers the system energy and opens an electronic band gap, the so-called Peierls gap \cite{Peierls}.
A similar transformation may occur in low-dimensional systems such as two-dimensional TMDs \cite{Yoffe}.
We suspect that the band gap opening that we observe at certain Li-ion site occupation fraction in the T' phase of MoS$_2$ and MoSe$_2$  is similar in nature to a Peierls gap opening. 

\begin{figure}[h!]
\centering
\subfloat[MoS$_2$]
{\includegraphics[width=0.43\textwidth]{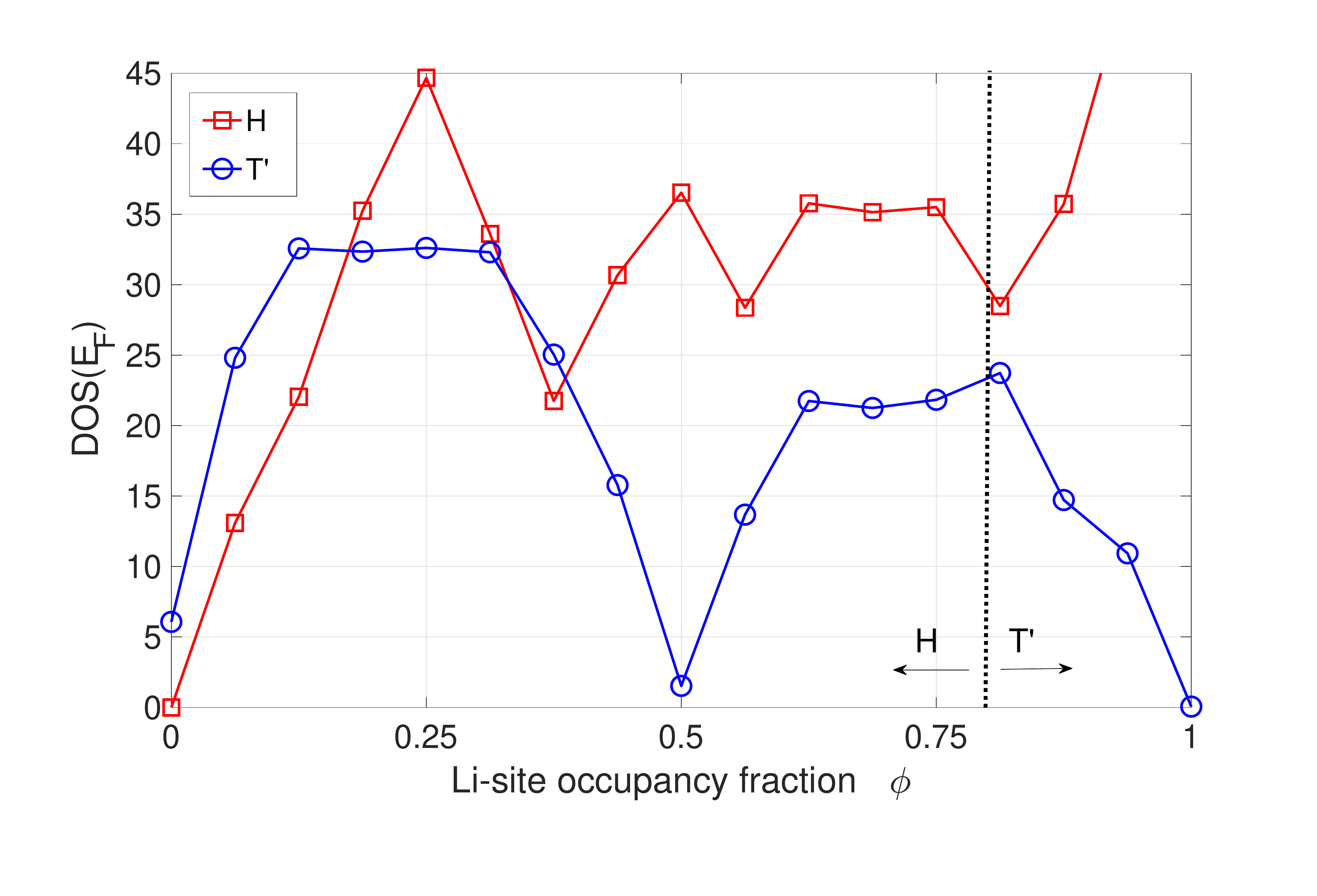}}\\
\subfloat[MoSe$_2$]
{\includegraphics[width=0.43\textwidth]{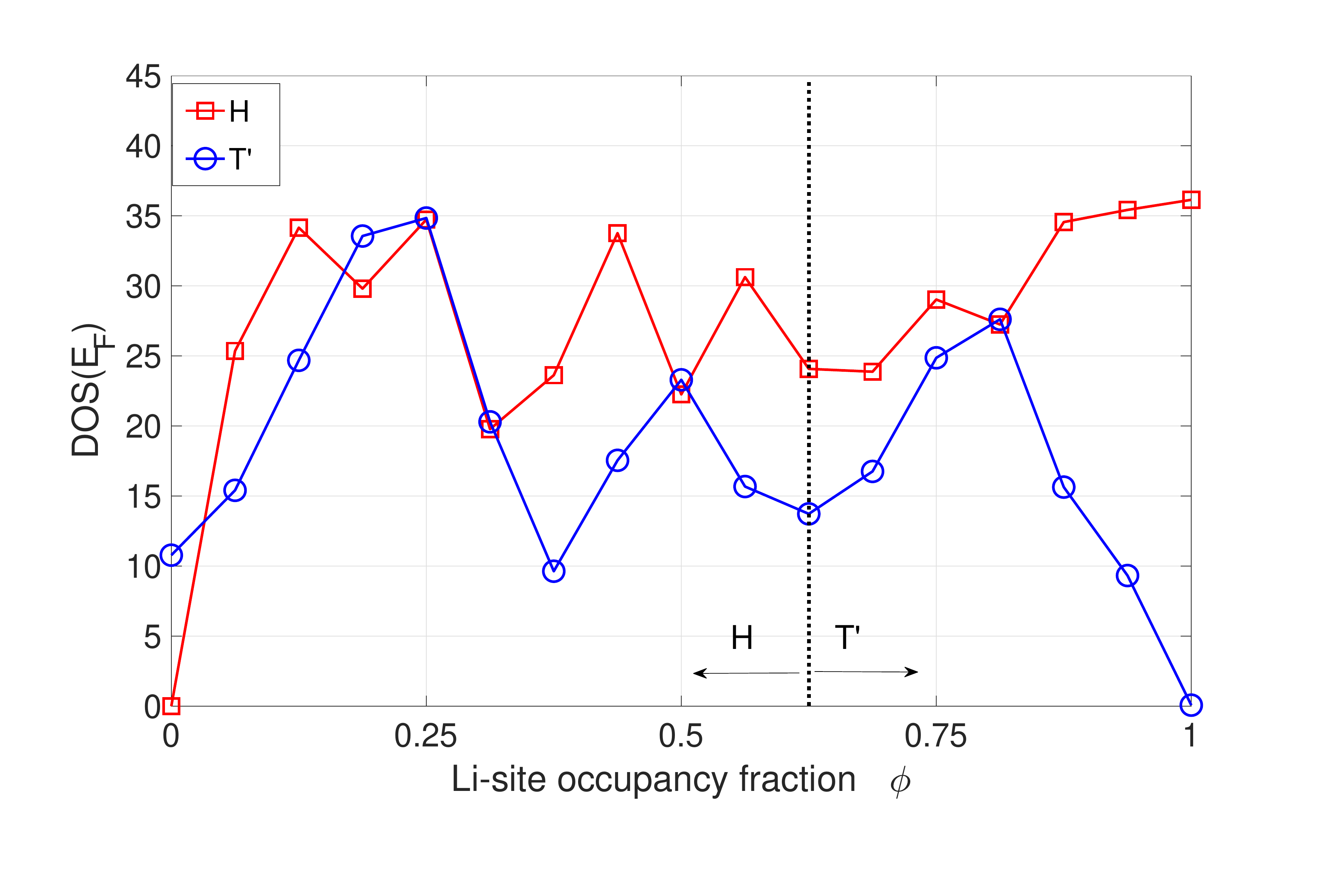}}\\
\subfloat[MoTe$_2$]
{\includegraphics[width=0.43\textwidth]{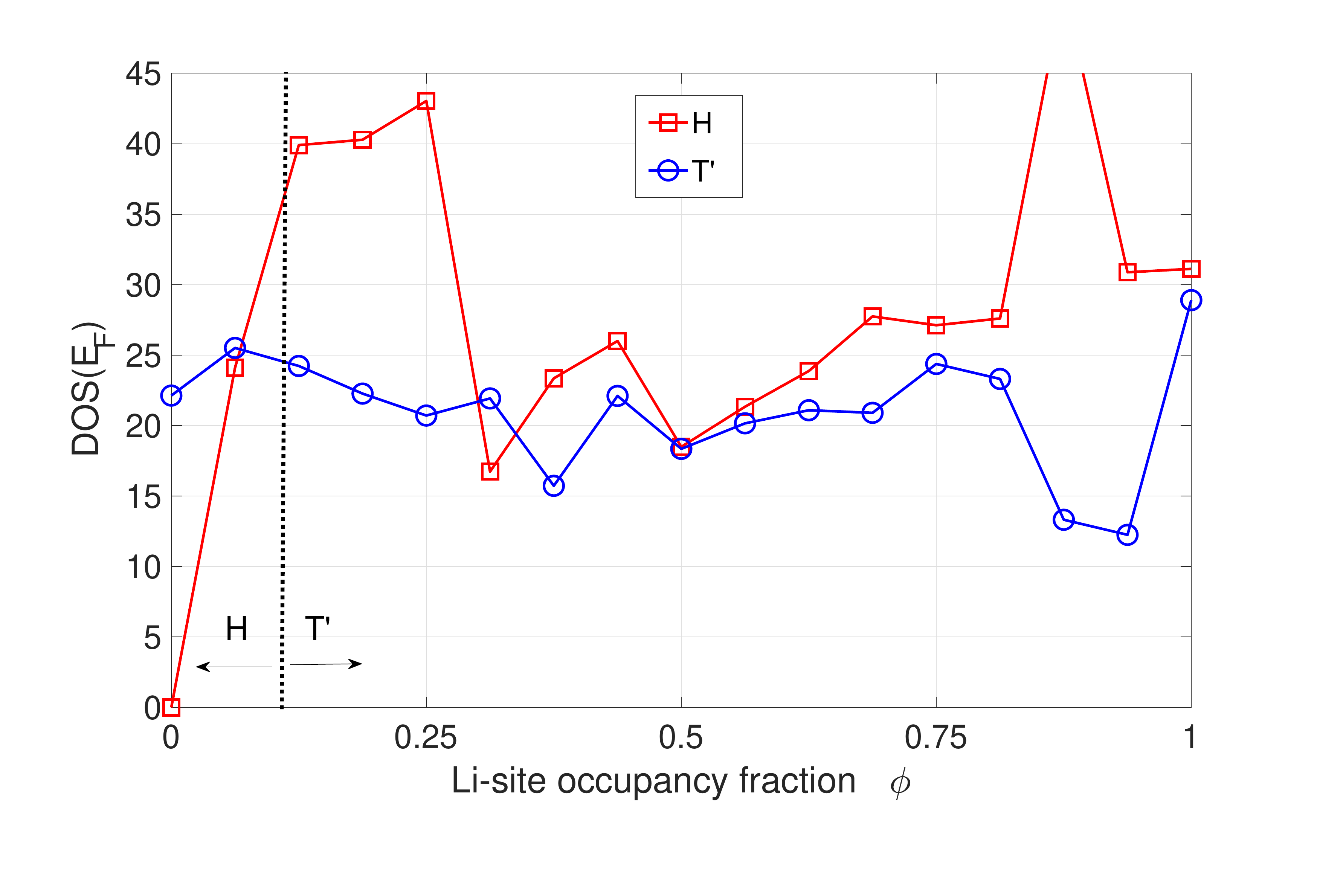}}
\caption{Density of states at the Fermi level obtained by statistical average within the NT ensemble for the two competing phases of Li ion intercalated bilayers: (a) MoS$_2$,  (b) MoSe$_2$, and  (c) MoTe$_2$. $T$=300~K. The dotted-dashed vertical lines indicates the ion concentrations where the systems transition form H to T' phase.}
\label{fig:dos_stat_avg}
\end{figure}

\fref{fig:dos_stat_avg} summarizes our results for the evolution of DOS(E$_F$) for each of the bilayer system/phase.
For MoSe$_2$ and MoTe$_2$ systems, DOS(E$_F$) increases more rapidly with increasing Li concentration for the H phase than for the T'. At relatively low ion concentration, the H phase may be therefore more conductive despite the fact that in the no-intercalation case it is semiconducting.
In general, as the concentration increases, the H phase remains the most conductive, with the peculiar situation in MoS$_2$ case where the T' phase seems to show very low conductivity near half Li sublattice occupancy ($x\approx 0.25$).
Importantly, as opposed to the expectation that a change in phase structure from H to T' correlates with a (positive) jump in electronic conductivity, we do not find this to hold for any of the three bilayer TMDs.
In fact our results suggest that as the system transitions between the two phases, the conductivity suffers a decrease instead of a (positive) jump.

\clearpage

\section{Conclusion}
We have studied the phase stability and electronic structure evolution  of Li-intercalated bilayer MoX$_2$ with X=S, Se or Te. 
Using first-principles calculations in combination with classical and machine learning modeling approaches, we find that H$\to$T' transition occurs at lower Li concentrations and with less change in free energy by increasing atomic mass of the chalcogen atom X.
While the electronic conductivity increases with increasing Li-ion concentration at low concentrations, we do not observe a (positive) conductivity jump at the H-T' phase transition point.
These findings have ramifications for ion insertion devices based on MoX$_2$ bilayers since two-dimensional layered materials such as TMDs could play an important role in the building blocks of neuromorphic/synaptic devices.
This work is a step toward understanding the fundamental mechanisms by which ion intercalation induces phase transition, which in turn affects the electronic properties. 
This knowledge is needed to control Li intercalation induced switching of conducting states in electrochemical devices.  

\section*{Acknowledgments}
This work was supported by the LDRD program at Sandia National Laboratories, and its support is gratefully acknowledged.
Sandia National Laboratories is a multimission laboratory managed and operated by National Technology and Engineering Solutions of Sandia, LLC., a wholly owned subsidiary of Honeywell International, Inc., for the U.S. Department of Energy's National Nuclear Security Administration under contract DE-NA0003525.
The views expressed in the article do not necessarily represent the views of the U.S. Department of Energy or the United States Government.


\appendix
\section*{Appendix A: Estimating finite size error} \label{sec:finite_size}
\renewcommand{\theequation}{A-\arabic{equation}} 
\renewcommand{\thefigure}{A-\arabic{figure}}
\setcounter{equation}{0}
\setcounter{subsection}{0}
\setcounter{figure}{0}

In this appendix we estimate the error in the free energy calculations due to using systems small enough for DFT simulation.
Expanding on the exposition of \sref{sec:free_energy}, the relative free energy $\Delta F$ is invariant to choice of reference $E^*$
\begin{align} 
\Delta F(N_\text{Li},T)
&=  k_B T \ln \left( \frac
{ \left( \sum_i \exp\left( -\frac{E_i-E^*}{k_B T} \right) \right)_\text{T'} }
{ \left( \sum_i \exp\left( -\frac{E_i-E^*}{k_B T} \right) \right)_\text{H} }
\right)
\end{align}
If we take $E^*$ to be the minimum energy of H and order the energies so that the first is the minimum and the last is the maximum, we obtain:
\begin{equation} 
\Delta F(N_\text{Li},T)  
= k_B T \ln \left( \frac 
{ \left( \exp\left( -\frac{E_1-E^*}{k_B T} \right) + \sum_{i=2}^N \exp\left( -\frac{E_i-E^*}{k_B T} \right) \right)_\text{T'} }
{ \left( 1 + \sum_{i=2}^N \exp\left( -\frac{E_i-E^*}{k_B T} \right) \right)_\text{H} }
\right)
\end{equation}
If the spacing between $E_i$ is much greater that $k_B T$ and the sums from 2 to $N$ are negligible, $\Delta F$ can be approximated by
\begin{equation} 
\Delta F(N_\text{Li},T)  = \min E_\text{T'} - \min E_\text{H}
\end{equation}
which motivates the determination of $\Delta F$ from  the lower convex hull of the sampled energies \ie minimum energies for each phase.

In a large system it is unlikely that the energies at fixed surface concentration will have large separations, so this approximation breaks down.
In this context the energies can be approximated by a continuous energy spectrum $p(E)$ so
\begin{equation} \label{eq:dF_cont}
\Delta F(N_\text{Li},T)  = k_B T \ln \left( \frac
{ \left( \int \exp\left( -\frac{E}{k_B T} \right) p(E) \mathrm{d}E \right)_\text{T'} }
{ \left( \int \exp\left( -\frac{E}{k_B T} \right) p(E) \mathrm{d}E \right)_\text{H} } \right)
\end{equation}
If we approximate $p(E)$ by normal distributions characterized by mean $\bar{E}$ and standard deviation $\sigma$ of our sampled energies, \eref{eq:dF_cont} reduces to
\begin{equation} \label{eq:corrected_E}
\Delta F(N_\text{Li},T)  
=
{ \left(-\bar{E} + \frac{\sigma^2}{2 k_B T}\right)_\text{T'}} -
{ \left(-\bar{E} + \frac{\sigma^2}{2 k_B T}\right)_\text{H}}
= - \Delta \bar{E}  + \Delta \left( \frac{\sigma}{2 k_B T} \sigma \right)
\end{equation}

\begin{figure}
\centering
{
\includegraphics[width=0.45\textwidth]{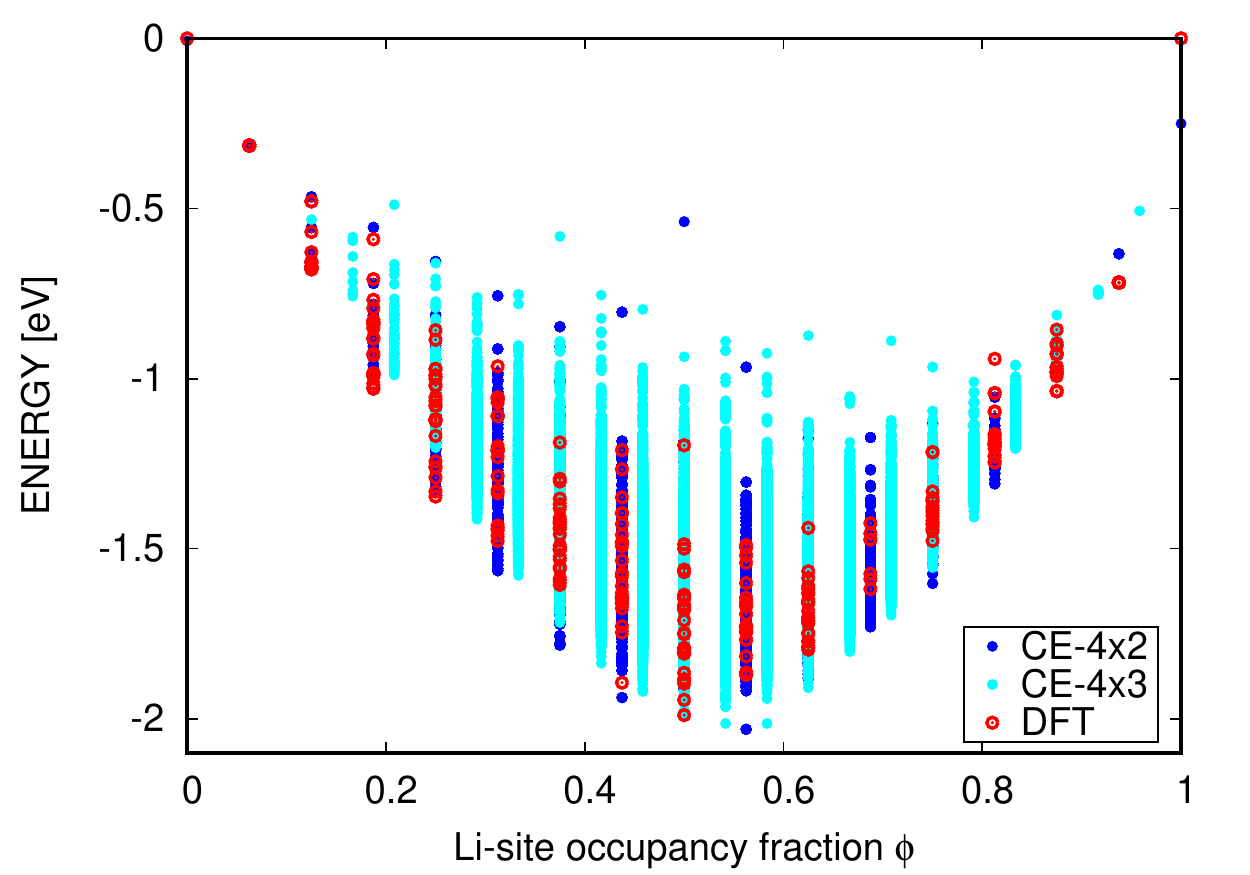}
\includegraphics[width=0.45\textwidth]{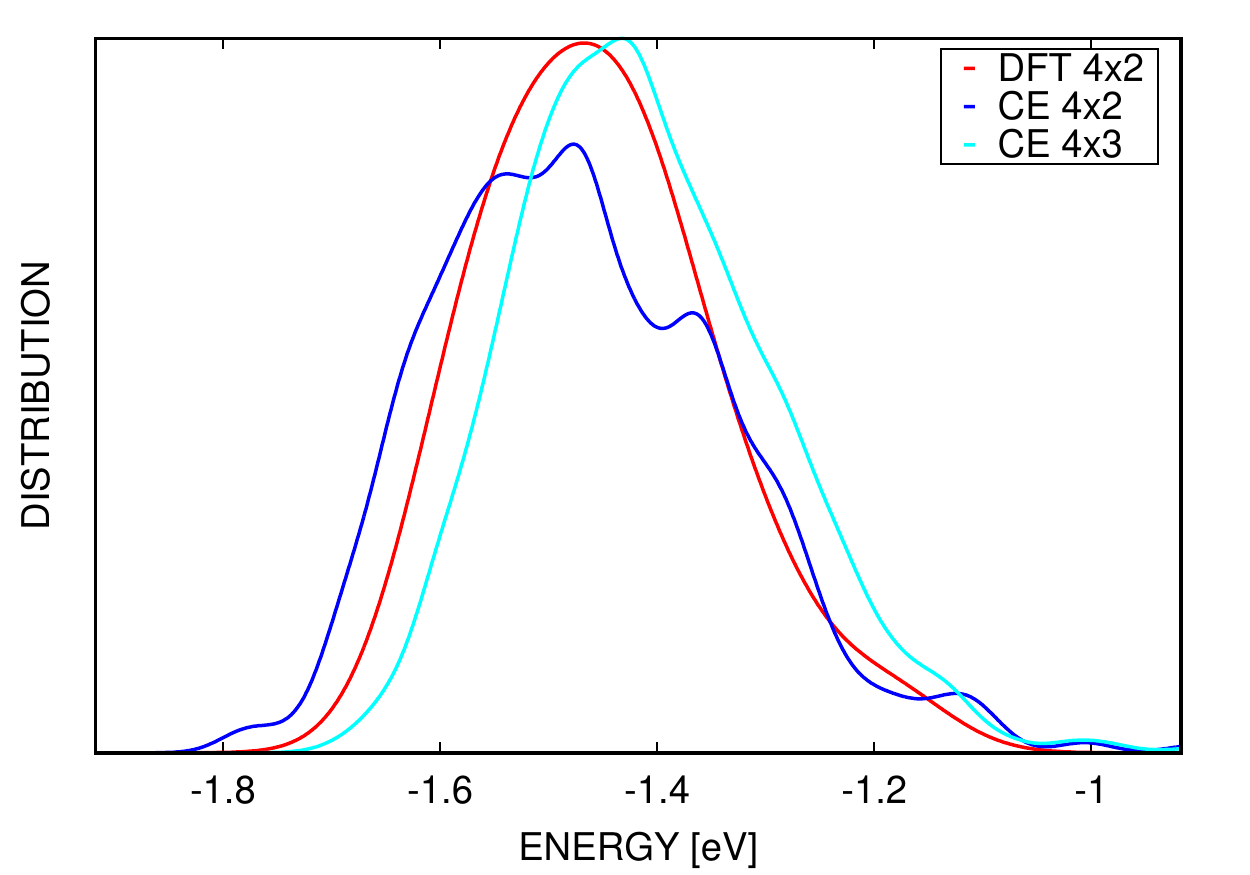}
}
\caption{Generalization of formation energy for T' phase to larger lattices using the CE order 2 model. (a) formation energy, and (b) energy distribution at fixed Li concentration.
}
\label{fig:generalization_pdfs}
\end{figure}

To explore the change in formation energy with larger systems we used the CE model to predict energies for all possible configurations of a 4$\times$3 system for T' MoS$_2$.
This system with 24 sites had a total of $2^{24}$=16,777,216 configurations versus $2^{16}$=65,536 for the 4$\times$2 16 site system.
As can be seen in \fref{fig:generalization_pdfs}a, the formation energies and their distribution change slightly despite the tremendous change in number of configurations.
In fact the larger lattice results appear to interpolate those of the smaller lattice and the distributions are similar, $\phi = $ 6/16 for 4$\times$2 and 9/24 for 4$\times$3 shown in \fref{fig:generalization_pdfs}b.

\begin{figure}[h!]
\centering
\subfloat[Energy distibution for T' calculated with CE order 2]
{ \includegraphics[width=0.45\textwidth]{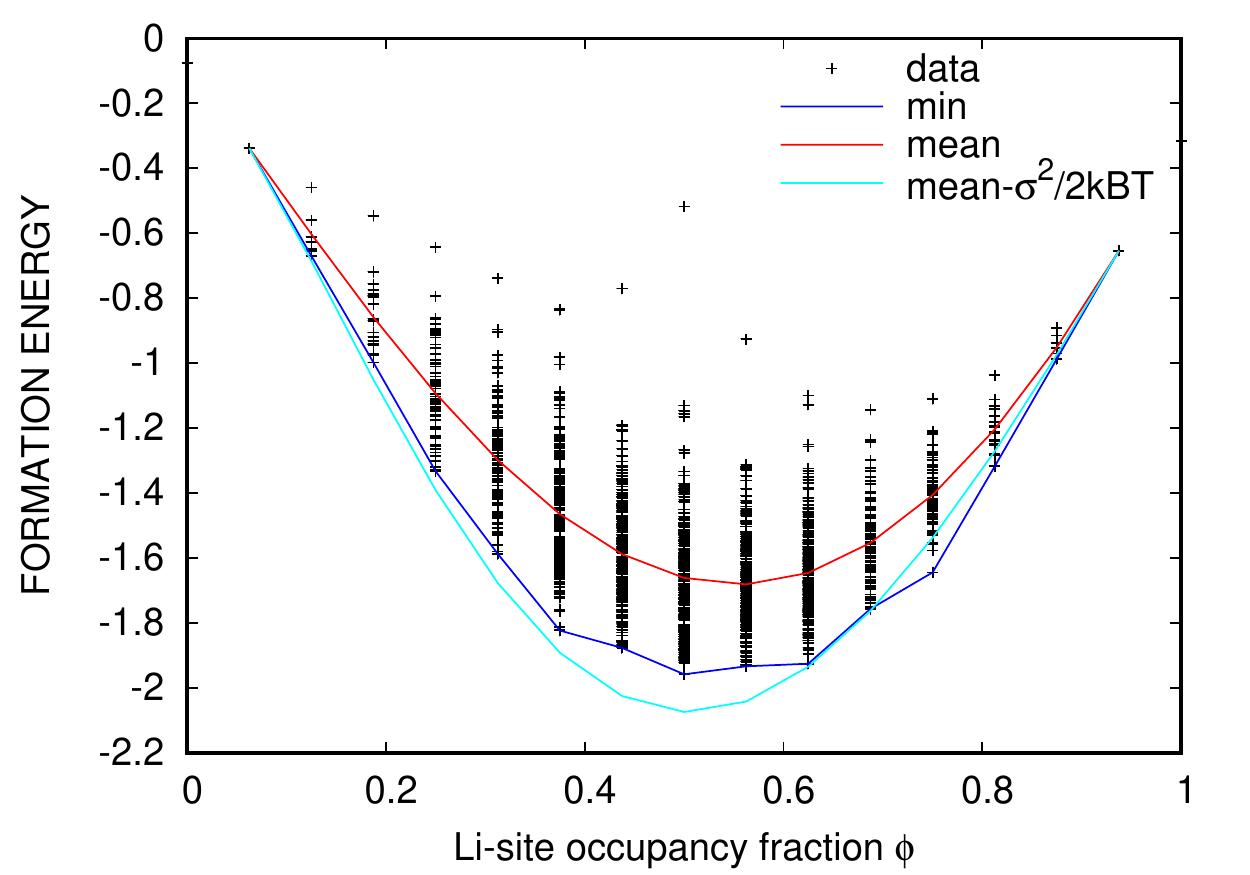} }
\subfloat[Standard deviation of the energy distributions]
{ \includegraphics[width=0.45\textwidth]{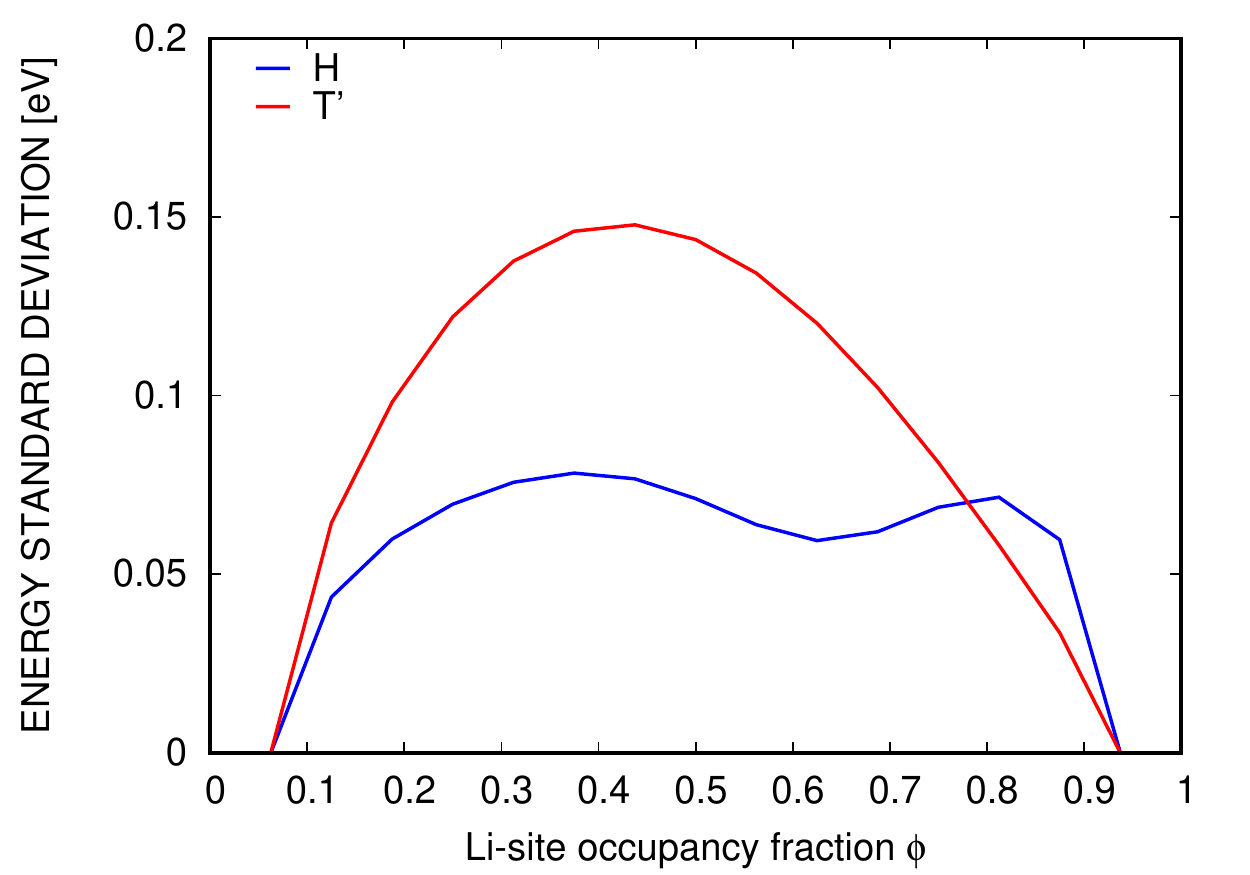} } 
\caption{Energy distribution correction: (a) comparison of energy corrected by the continuous normal distribution formula and the lower convex hull of the samples, (b) standard deviation estimates for the two phases obtained directly from the data.
}
\label{fig:corrected_energy}
\end{figure}

Motivated by this observation, we also used the continuous distribution formula \eref{eq:corrected_E} to estimate the errors from using a small sample of the configuration space.
\fref{fig:corrected_energy}a shows the estimate from \eref{eq:corrected_E} is comparable to the lower convex hull obtained from the minimum energy as a function of $\phi$. 
Here the energy distributions means and standard deviations were obtained from samples generated by the CE model.
The standard deviations of the energy distributions as a function of site occupancy $\phi$ shown in \fref{fig:corrected_energy}b indicate that they are small and comparable for the two phases and do not affect the phase transitions predicted from the minimum energies.

\appendix
\section*{Appendix B: Surrogate model comparison}
\renewcommand{\theequation}{B-\arabic{equation}} 
\renewcommand{\thefigure}{B-\arabic{figure}}
\setcounter{equation}{0}
\setcounter{subsection}{0}
\setcounter{figure}{0}

We first compared several surrogate models (CE, MLP,  CGCNN and transfer learning) using a less computationally intensive (but artificial) system, namely an unrelaxed MoS$_2$ monolayer in the H form.
We have generated $2550$ configurations with Li ion randomly adsorbed on one face of the monolayer, with the possible adsorption sites forming a 16-site Li sublattice.
The DFT calculations are carried non-self-consistently, hence they are fast computationally but the output energies do not have physical significance.

Within the standard CE method ($n_{CE}=1$) the energies of monolayer, MoS$_2$ are predicted  with relatively good accuracy, namely we achieve a mean average error (MAE) of $36$ meV.
However, we find a significant improvement by increasing $n_{CE}$ to $2$, in which case the MAE decreases to $7$ meV.
A similar trend is obtained via the machine learning techniques.
Within CGCNN it is straightforward to obtain a MAE of $20$ meV and a similar prediction error is obtained  within the transfer learning technique upon fine-tuning.
We note that while the machine learning techniques yield better prediction accuracy than the standard ($n_{CE}=1$) CE method, a higher-order polynomial expansion within CE (which is still computationally very efficient) results in a significantly smaller MAE than CGCNN and transfer learning CNN.
 
\begin{figure}[h!]
\centering
\subfloat[CE]
{\includegraphics[width=0.5\textwidth]{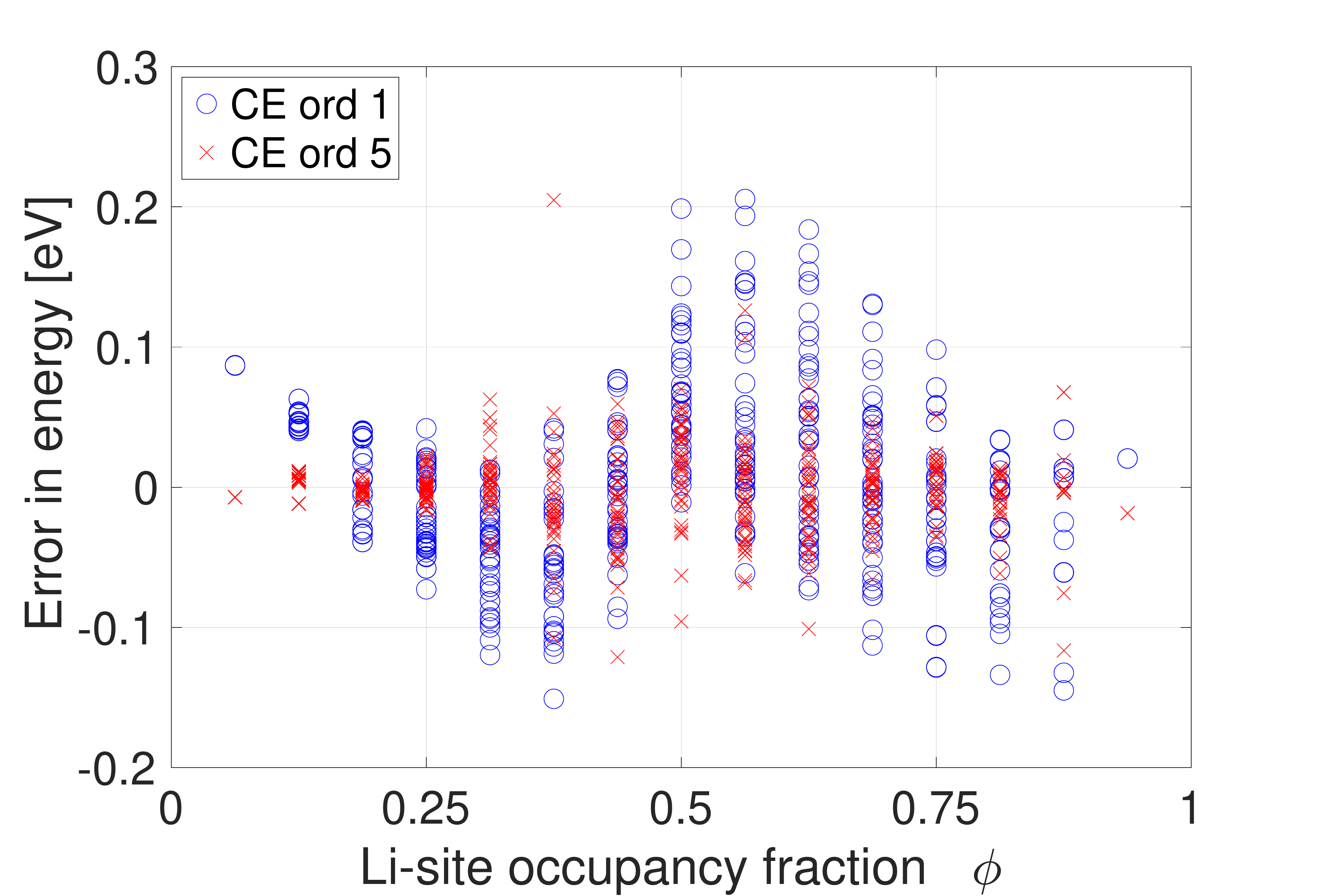}}
\subfloat[NN]
{\includegraphics[width=0.5\textwidth]{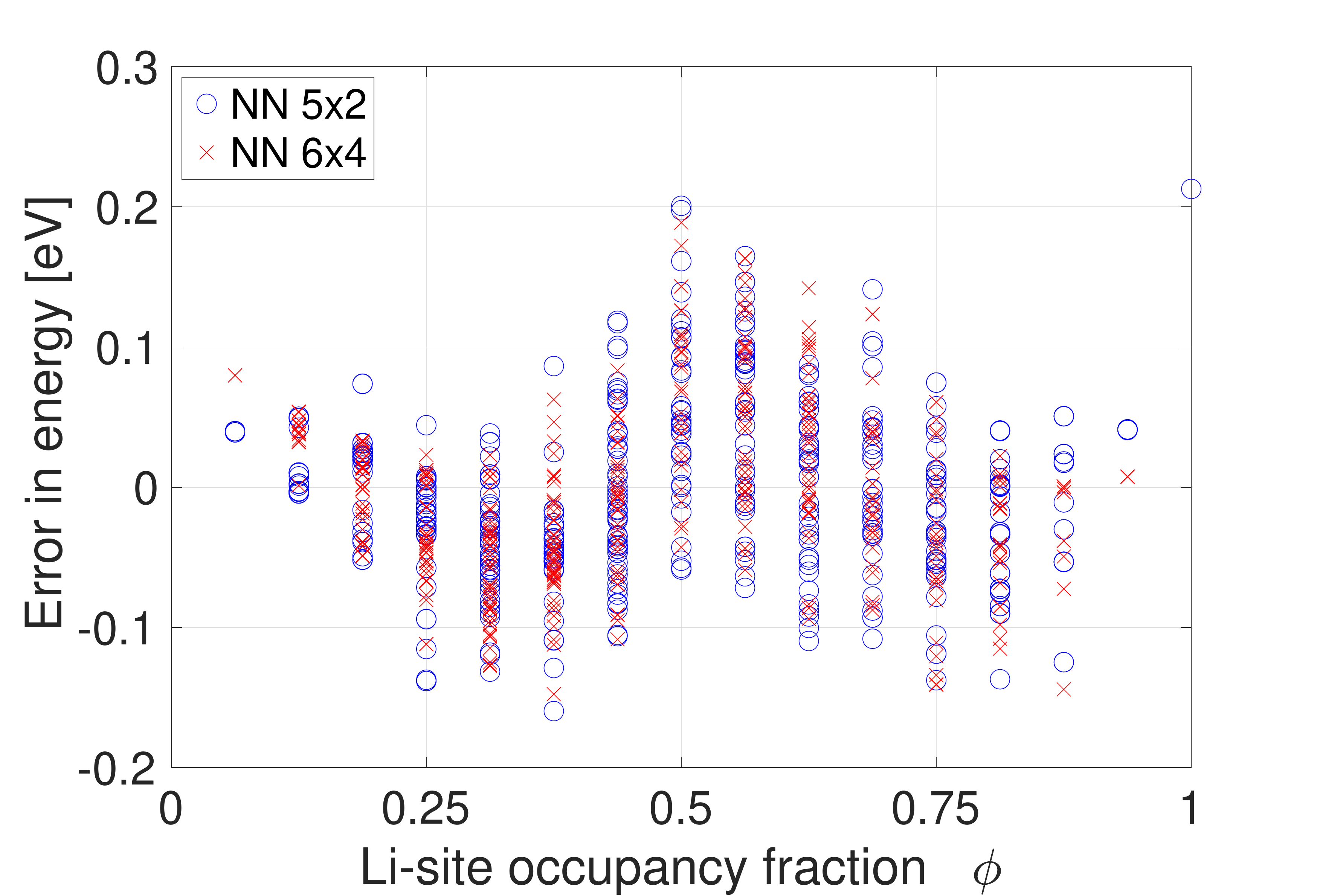}}
\caption{Formation energy errors for CE and NN models for 2500 samples of the T' phase of MoS$_2$. Note for the NN model $N_1\times N_2$ denotes an MLP with N$_1$ nodes in the first layer and $N_2$ nodes in the second.
}
\label{fig:energy_error}
\end{figure}

We have also compared CE and machine learning techniques using a realistic system, namely Li ion intercalated bilayer MoS$_2$ (T' phase) simulated within self-consistent DFT that included relaxation of forces and stress for each of the 2500 randomly generated configurations.
The surrogates were trained on an $80\%$ subset of the data and tested with 5-fold validation.
\fref{fig:energy_error} shows that the errors decrease with the order of the polynomial expansion for the CE model and the size of the network for the MLP NN model using the same input features.
The standard (1st order) CE method yields a MAE of 46 meV while 5th order CE yields the best MAE (20 meV). 
We find that the 5$\times$2, 6$\times$2 and 6$\times$4 NN networks have similar MAE (47 meV).
Both formulations showed little bias in representing the formation energy.

\begin{figure}[h!]
\centering
{\includegraphics[width=0.6\textwidth]{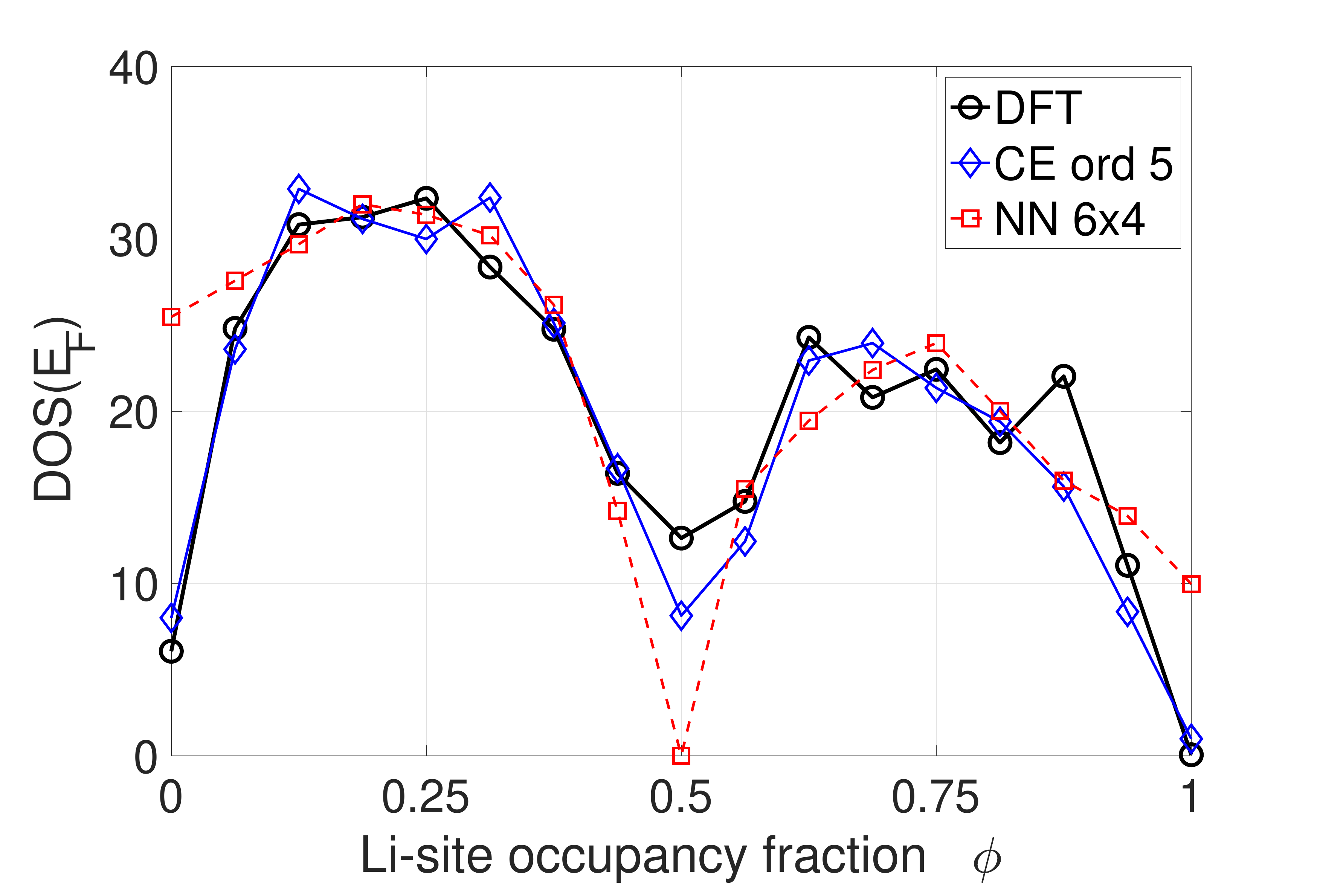}}
\caption{Predicted (statistically averaged) density of states at the Fermi level of the T' phase of MoS$_2$ for CE and NN models compared to DFT. CE and NN are trained using the 2500 DFT configurations.}
\label{fig:dos_error}
\end{figure}

\fref{fig:dos_error} shows the predictions in corresponding representations of the DOS (for an energy near the Fermi level) data. 
Again the errors are controllable but in order to capture details such as the decrease of DOS(E$_F$) for Li-site occupancy fraction $\phi\approx 0.5$ 
 they require a larger order (n$_{CE}=5$) polynomial expansion for CE and a larger size (6$\times$4) of the network for NN.
The two formulations show distinct bias but produce models with comparable errors.

\begin{thebibliography}{10}

\bibitem{murray2018basic}
Cherry Murray, Supratik Guha, Dan Reed, Gil Herrera, Kerstin Kleese~van Dam,
  Sayeef Salahuddin, James Ang, Thomas Conte, Debdeep Jena, Robert Kaplar,
  et~al.
\newblock Basic research needs for microelectronics: Report of the office of
  science workshop on basic research needs for microelectronics, october
  23--25, 2018.
\newblock Technical report, USDOE Office of Science (SC)(United States), 2018.

\bibitem{fuller2017li}
Elliot~J Fuller, Farid~El Gabaly, Fran{\c{c}}ois L{\'e}onard, Sapan Agarwal,
  Steven~J Plimpton, Robin~B Jacobs-Gedrim, Conrad~D James, Matthew~J
  Marinella, and A~Alec Talin.
\newblock Li-ion synaptic transistor for low power analog computing.
\newblock {\em Advanced Materials}, 29(4):1604310, 2017.

\bibitem{fuller2019parallel}
Elliot~J Fuller, Scott~T Keene, Armantas Melianas, Zhongrui Wang, Sapan
  Agarwal, Yiyang Li, Yaakov Tuchman, Conrad~D James, Matthew~J Marinella,
  J~Joshua Yang, et~al.
\newblock Parallel programming of an ionic floating-gate memory array for
  scalable neuromorphic computing.
\newblock {\em Science}, 364(6440):570--574, 2019.

\bibitem{yang2018all}
Chuan-Sen Yang, Da-Shan Shang, Nan Liu, Elliot~J Fuller, Sapan Agrawal, A~Alec
  Talin, Yong-Qing Li, Bao-Gen Shen, and Young Sun.
\newblock All-solid-state synaptic transistor with ultralow conductance for
  neuromorphic computing.
\newblock {\em Advanced Functional Materials}, 28(42):1804170, 2018.

\bibitem{nadkarni2019modeling}
Neel Nadkarni, Tingtao Zhou, Dimitrios Fraggedakis, Tao Gao, and Martin~Z
  Bazant.
\newblock Modeling the metal--insulator phase transition in lixcoo2 for energy
  and information storage.
\newblock {\em Advanced Functional Materials}, 29(40):1902821, 2019.

\bibitem{enyashin2013line}
Andrey~N Enyashin, Maya Bar-Sadan, Lothar Houben, and Gotthard Seifert.
\newblock Line defects in molybdenum disulfide layers.
\newblock {\em The Journal of Physical Chemistry C}, 117(20):10842--10848,
  2013.

\bibitem{benavente2002intercalation}
E~Benavente, MA~Santa~Ana, F~Mendiz{\'a}bal, and G~Gonz{\'a}lez.
\newblock Intercalation chemistry of molybdenum disulfide.
\newblock {\em Coordination chemistry reviews}, 224(1-2):87--109, 2002.

\bibitem{Zhao}
Wei Zhao, Fuqiang Huang, Jie Pan, Yuqiang Fang, Xiangli Che, Dong Wang, Kejun
  Bu, and Fuqian Huang.
\newblock Metastable mos$_2$: Crystal structure, electronic band structure,
  synthetic approach and intriguing physical properties.
\newblock {\em Chemistry European Journal Review}, 24:15942, 2018.

\bibitem{sun2016origin}
Xiaoli Sun, Zhiguo Wang, Zhijie Li, and Yong~Qing Fu.
\newblock Origin of structural transformation in mono-and bi-layered molybdenum
  disulfide.
\newblock {\em Scientific reports}, 6(1):1--9, 2016.

\bibitem{zhang2018reversible}
Jinsong Zhang, Ankun Yang, Xi~Wu, Jorik van~de Groep, Peizhe Tang, Shaorui Li,
  Bofei Liu, Feifei Shi, Jiayu Wan, Qitong Li, et~al.
\newblock Reversible and selective ion intercalation through the top surface of
  few-layer mos 2.
\newblock {\em Nature communications}, 9(1):1--9, 2018.

\bibitem{pandey2016phase}
Mohnish Pandey, Pallavi Bothra, and Swapan~K Pati.
\newblock Phase transition of mos2 bilayer structures.
\newblock {\em The Journal of Physical Chemistry C}, 120(7):3776--3780, 2016.

\bibitem{lu2020lithium}
Zheyu Lu, Stephen Carr, Daniel~T Larson, and Efthimios Kaxiras.
\newblock Lithium intercalation in {MoS}$_2$ bilayers and implications for
  {M}oir\'e flat bands.
\newblock {\em Physical Review B}, 2020.

\bibitem{li2016ferroelasticity}
Wenbin Li and Ju~Li.
\newblock Ferroelasticity and domain physics in two-dimensional transition
  metal dichalcogenide monolayers.
\newblock {\em Nature communications}, 7(1):1--8, 2016.

\bibitem{van1998first}
Anton Van~der Ven, MK~Aydinol, G~Ceder, Georg Kresse, and Jurgen Hafner.
\newblock First-principles investigation of phase stability in li x coo 2.
\newblock {\em Physical Review B}, 58(6):2975, 1998.

\bibitem{urban2016computational}
Alexander Urban, Dong-Hwa Seo, and Gerbrand Ceder.
\newblock Computational understanding of li-ion batteries.
\newblock {\em npj Computational Materials}, 2(1):1--13, 2016.

\bibitem{kresse1996efficiency}
Georg Kresse and J{\"u}rgen Furthm{\"u}ller.
\newblock Efficiency of ab-initio total energy calculations for metals and
  semiconductors using a plane-wave basis set.
\newblock {\em Computational materials science}, 6(1):15--50, 1996.

\bibitem{blochl1994projector}
Peter~E Bl{\"o}chl.
\newblock Projector augmented-wave method.
\newblock {\em Physical review B}, 50(24):17953, 1994.

\bibitem{perdew1996generalized}
John~P Perdew, Kieron Burke, and Matthias Ernzerhof.
\newblock Generalized gradient approximation made simple.
\newblock {\em Physical review letters}, 77(18):3865, 1996.

\bibitem{he2014}
Jiangang He, Kerstin Hummer, and Cesare Franchini.
\newblock Stacking effects on the electronic and optical properties of bilayer
  transition metal dichalcogenides mos$_2$, mose$_2$, ws$_2$, and wse$_2$.
\newblock {\em Physical Review B}, 89:075409, 2014.

\bibitem{sanchez1984generalized}
Juan~M Sanchez, Francois Ducastelle, and Denis Gratias.
\newblock Generalized cluster description of multicomponent systems.
\newblock {\em Physica A: Statistical Mechanics and its Applications},
  128(1-2):334--350, 1984.

\bibitem{de1994cluster}
Didier De~Fontaine.
\newblock Cluster approach to order-disorder transformations in alloys.
\newblock {\em Solid state physics}, 47:33--176, 1994.

\bibitem{li1994lattice}
W~Li, JN~Reimers, and JR~Dahn.
\newblock Lattice-gas-model approach to understanding the structures of lithium
  transition-metal oxides lim o 2.
\newblock {\em Physical Review B}, 49(2):826, 1994.

\bibitem{xie2018crystal}
Tian Xie and Jeffrey~C Grossman.
\newblock Crystal graph convolutional neural networks for an accurate and
  interpretable prediction of material properties.
\newblock {\em Physical review letters}, 120(14):145301, 2018.

\bibitem{Matlab}
MATLAB.
\newblock {\em version 9.9.0 (R2020b)}.
\newblock The MathWorks Inc., Natick, Massachusetts, 2020.

\bibitem{szegedy2015going}
Christian Szegedy, Wei Liu, Yangqing Jia, Pierre Sermanet, Scott Reed, Dragomir
  Anguelov, Dumitru Erhan, Vincent Vanhoucke, and Andrew Rabinovich.
\newblock Going deeper with convolutions.
\newblock In {\em Proceedings of the IEEE conference on computer vision and
  pattern recognition}, pages 1--9, 2015.

\bibitem{doping2017}
Erica~E. Hroblak, Alessandro Principi, Hui Zhao, and Giovanni Vignale.
\newblock Electrically induced charge-density waves in a two-dimensional
  electron liquid: Effects of negative electronic compressibility.
\newblock {\em Physical Review B}, 96:075422, 2017.

\bibitem{cdw1974}
J.A. Wilson, F.~J. Di~Salvo, and Mahajan S.
\newblock Charge-density waves in metallic, layered, transition-metal
  dichalcogenides.
\newblock {\em Physical Review Letters}, 32:882, 1974.

\bibitem{Peierls}
Rudolf~Ernst Peierls.
\newblock {\em Quantum theory of solids}.
\newblock Clarendon Press, 1996.

\bibitem{Yoffe}
RH~Friend and AD~Yoffe.
\newblock Electronic properties of intercalation complexes of the transition
  metal dichalcogenides.
\newblock {\em Advances in Physics}, 36(1):1--94, 1987.

\end{thebibliography}
\end{document}